\documentclass[review]{elsarticle}
\usepackage{subfigure}
\usepackage{indentfirst}
\usepackage{bm}
\usepackage{threeparttable}
\usepackage{multirow}
\usepackage{amsmath}
\usepackage{array}
\usepackage{amssymb}
\usepackage{graphicx}
\usepackage{url}
\usepackage{hyperref}
\usepackage{epstopdf}
\usepackage{booktabs}
\usepackage{color}

\journal{Computers \& Mathematics with Applications}

\begin{document}

\begin{frontmatter}

\title{A simplified finite volume lattice Boltzmann method for simulations of fluid flows from laminar to turbulent regime, Part \uppercase\expandafter{\romannumeral1}: Numerical framework and its application to laminar flow simulation}
\author{Yong Wang$^a$}
\ead{wangyong19890513@mail.nwpu.edu.cn}

\author[]{Chengwen Zhong$^a$\corref{mycorrespondingauthor}}
\ead{zhongcw@nwpu.edu.cn}
\author[]{Jun Cao$^b$\corref{mycorrespondingauthor}}
\cortext[mycorrespondingauthor]{Corresponding author}
\ead{jcao@ryerson.ca}

\author[]{Congshan Zhuo$^a$}
\ead{zhuocs@nwpu.edu.cn}
\address{$^a$National Key Laboratory of Science and Technology on Aerodynamic Design and Research, Northwestern Polytechnical University, Xi'an, Shaanxi 710072, China \\
$^b$Department of Mechanical \& Industrial Engineering, Ryerson University, Toronto, Ontario, Canada M5B 2K3}

\begin{abstract}
In this paper, a finite volume lattice Boltzmann method (FVLBM) based on cell-center unstructured girds is presented and full studied to simulate the incompressible laminar flows, which is simple modified from the cell-vertex unstructured girds FVLBM proposed by Stiebler et al. [Computers \& Fluids, 2006, 35(8): 814-819]. Compared with other complex flux reconstruct methods, the computational cost of present scheme is little and can achieve second-order spatial accuracy, the temporal accuracy is adjustable depending on the temporal discretized methods. Different boundary conditions are illustrated and easy implement to the complex geometries. Four cases are testified to validate the present method, including one plate driven Couette flow for accuracy test, flow in the square cavity, flow over the single circular cylinder and more complex double circular cylinders. Numerical experiments show that the present scheme can use relatively few grid cells to simulate relatively higher Reynolds number flow, steady and unsteady flows, demonstrate the good capability of the present method.
\end{abstract}


\end{frontmatter}

\section{Introduction}
Starting from the appearance of lattice Boltzmann method (LBM)\cite{mcnamara1988use}, after nearly three decades of development, LBM has become a relatively mature method and can be considered as an alternative to the macro-method based on the Navier-Stokes (NS) equations to some degree. Nowadays, LBM has a wide application than macro-method as its distinctive feature\cite{succi2001lattice}\cite{guo2013lattice} and will attract more and more attention in the future\cite{succi2015lattice}. As originated from lattice-gas automata (LGA)\cite{wolf2000lattice}, the standard LBM is usually based on Cartesian grid and can be split into streaming and collision steps. Using the Chapman-Enskog perturbation expansion, both spatial and temporal accuracy of LBM can achieve second-order\cite{guo2013lattice}. Though the standard LBM is easy to implement, the uniform Cartesian grid and unadjustable temporal and spatial accuracy become its the shortcomings, which limit its scope of applications. For the complex geometries, the stepped polygons are used to fit the border as the restriction of the Cartesian grid, so the fitting accuracy is lower compared with body-fitted grid. Besides, in two-dimensional (2D) computational domain, the square elements of grid have the identical scales in two directions, so to refine the grid near the wall boundary to enhance the accuracy of fitting will increase the amount of grid cells extremely. To simulate the high Reynolds number flows, as the change of physical quantities in the normal direction to the wall are much larger than the streamwise direction, the body-fitted grid usually used in macro-method to reduce the computational cost. At the same Reynolds number, the amount of girds is unacceptable if Cartesian grid is used. Besides, the fixed accuracy in spatial and temporal may has difficult to some problems that higher order spatial and temporal scheme is needed.

To remove the limit of Cartesian grid, some methods using non-uniform grid have been developed. One way is some modifications are added into the standard LBM. The widely used method is grid-refinement and multi-block methods which first introduced by Filippova et al.\cite{filippova1998grid}. Some improvements can be found in Ref.~\cite{dupuis2003theory,guo2003domain}. In these methods, all of the blocks are still use the Cartesian grid and implement the standard LBM, but the grid resolution of each block can be changed depending on the variation of physical qualities. During the calculation, the unknown distribution functions at the interfaces of blocks need to interpolate. The main problem of this method is that if the unreasonable interpolation schemes are used, the physical qualities may discontinuous at the interfaces of blocks, and the disturbances may transport in the interior of each blocks, finally impact the results of calculation. In addition, for high Reynolds number flow simulations, though the multi-block techniques can decline the amount of grid cells at far-field, the number of grid nodes at near wall domain is still large to be accepted. One of other important methods is the interpolation supplemented LBE (ISLBE)\cite{he1997lattice} which the nonuniform and body-fitted grid can be used. The main problems of these methods are discussed in Ref.~\cite{lallemand2000theory}.

Another way to improve the LBM is to abandon the streaming and collision steps in standard LBM completely. From theoretical analysis in Ref.~\cite{he1997theory}, we can find that the standard LBM is a special finite difference method (FDM) to solve the Boltzmann BGK equation. So, the well-developed finite difference method, finite volume method (FVM) and finite element method (FEM) used in macro-method may also can be used to LBM. Ref.~\cite{guo2003explicit,nannelli1992lattice,peng1998lattice,lee2001characteristic} are some works about these methods.

For the finite volume LBM (FVLBM), the studies can be divided into two paths as there exist two different types of grid: structured grid and unstructured grid. In addition, the grid can be divided into two sub-types: cell-vertex type and cell-center type. All of these types are used in FVLBM, and more detail information about grid can be found in Ref.~\cite{patil2009finite}. As a consequence, the reconstruction of distribution function at the cell interfaces are different depended on the grid types. The representative works for the structured grid FVLBM can be found in Ref \cite{razavi2009flux,choi2010simple}. In the formulation of unstructured grid FVLBM, the most simple reconstruct method is center scheme\cite{peng1998lattice,ubertini2004recent,rossi2005unstructured}. Though the spatial accuracy can achieve second-order,
in high Reynolds number flow simulation, this finite volume lattice Boltzmann method is unstable. It will appear numerical instability to simulate higher Reynolds number flows\cite{stiebler2006upwind}. So the test cases in these studies are restricted to low Reynolds number flows. To overcome the drawback of center scheme on unstructured grid, the upwind schemes are introduced into FVLBM\cite{stiebler2006upwind,patil2009finite,lifinite} and much higher Reynolds number flows are simulated to validate the methods. Patil et al. introduce the Roe's flux-difference splitting scheme and the limiters to the unstructured grid\cite{patil2009finite}. The main problem of this work is that the numerical diffusion is larger, which causes the results are not very good for high Reynolds number flows. Li et al. present a modified Roe's scheme based on the formulation of Patil et al.'s and have good results compared with the latter's\cite{lifinite}. Another representative work presented by Chen et al. is using the Lax-Wendroff scheme to reconstruct the distribution function at cell interfaces. However, the performance of this scheme at higher Reynolds number flows need further studies\cite{chen2015unified}. Besides, compared with the adequate researches about the performance of FVLBM for simulating the steady flow on unstructured grid, the studies for unsteady flow are insufficient. Recently, Patil et al. present an improved scheme for unsteady flow and have more accuracy results compared with their originated work\cite{patil2012two}. The main defect of this work is that the stencil for calculation the gradient of distribution function is not compact, so the computational efficiency is lower for parallel computing. In the numerical simulation, boundary conditions play an important role in fluid dynamics since they are essential in the determination of the solution of the flow\cite{guo2013lattice}, so new problems are emerged for FVLBM as methods developed in standard LBM can not transplant to FVLBM directly. The work on this topic is scanty\cite{peng1999boundary,chen2015unified} and need further studies. The relationship between the viscosity of fluid and the relaxation time are the reflection of flow features. The detail analysis about this relationship at different temporal and spatial discretized schemes through Chapman-Enskog expansion\cite{patil2013chapman,misztal2015detailed} are also scarce and need further research.

The main objective of this paper is to study the performance of unstructured grid FVLBM used in this work for more higher Reynolds number flows, steady and unsteady flows. Since the upwind scheme presented by Stiebler et al.\cite{stiebler2006upwind} is little computational cost compared with other upwind schemes used in unstructured grid FVLBM, it is attractive if a modified formulation based on this simple scheme coupling with more easy implement of boundary conditions can apply to extended computational scope.

The rest of this paper is organized as flows. Section \ref{sec:Numerical Method} presents the numerical method used in this study, including the reconstruct scheme for the distribution functions at cell interface, temporal discrete formulation and boundary conditions. Several test cases are conducted to validate the method used in this paper in Section \ref{Sec:Numerical experiments}. Finally. the conclusion will be grouped in Section \ref{Sec:conclusions}.
\section{The Numerical Method}\label{sec:Numerical Method}

\subsection{D2Q9 version of lattice Boltzmann method}
To solve the Boltzmann equation, the most popular method for LBM is to discretize the equation in velocity space with finite and regular discrete velocities, and to model the collision term with Bhatnagar-Gross-Krook approximation. Eq. \eqref{bgkeq} shows the discrete Boltzmann equations (DBE):
\begin{equation}\label{bgkeq}
     \frac{\partial{f_{\alpha}(\bm{x},t)}}{\partial{t}} + \bm{e}_{\alpha} \cdot \nabla f_{\alpha}(\bm{x},t)
  = -\frac{1}{\tau}[f_{\alpha}(\bm{x},t) - f_{\alpha}^{eq}(\bm{x},t)],
\end{equation}
where $f_\alpha(\bm{x},t)$ is the particle distribution function at location $\bm{x}$ and time $t$, subscript $\alpha$ is the $\alpha^{th}$ discrete velocity direction and $\bm{e}_{\alpha}$ is corresponding discrete velocity vector,  $f_{\alpha}^{eq}$ is the equilibrium distribution function, and $\tau$ is the relaxation time. Repeated subscript $\alpha$ do not imply Einstein summation. For discrete velocity model, in this paper, the D2Q9 lattice model\cite{qian1992lattice} is used and the formulation is
\begin{equation}
\bm{e}_\alpha = c\begin{cases}
                 \left(0,0\right)  & for \quad \alpha = 0, \\
                 \left(\cos[(\alpha -1)\pi/2],\sin[(\alpha -1)\pi/2]\right) & for \quad \alpha = 1,2,3,4, \\
                 \left(\sqrt{2}\cos[(2\alpha -1)\pi/4],\sqrt{2}\sin[(2\alpha -1)\pi/4]\right) & for \quad \alpha = 5,6,7,8,
                 \end{cases}
\end{equation}
where $c$ is an arbitrary constant related to the speed of lattice sound $c_s$($c_s=1/\sqrt{3}$) and is given by $c=c_s\sqrt{3}$. Here, the $c$ is equal to 1.

The equilibrium distribution functions for D2Q9 is
\begin{equation}\label{feq}
f_{\alpha}^{eq}(\rho, \bm{u})=\rho\omega_{\alpha}[1 + \frac{\bm{e}_{\alpha} \cdot \bm{u}}{c_s^2}
                                + \frac{(\bm{e}_{\alpha} \cdot \bm{u})^2}{2c_s^4}
                                - \frac{u^2}{2c_s^2}],
\end{equation}
where $\rho$ and $\bm{u}$ are the density and the velocity of fluid, respectively. $\omega_\alpha$ is the weighting factor, the D2Q9 model is
\begin{equation}
\omega_\alpha =  \begin{cases}
                 4/9  & for \quad \alpha = 0,       \\
                 1/9  & for \quad \alpha = 1,2,3,4, \\
                 1/36 & for \quad \alpha = 5,6,7,8.
                 \end{cases}
\end{equation}

The relationship between macro physical variables and distribution functions is
\begin{equation}
       \rho=\sum_{\alpha=0}^{8}f_{\alpha},   \quad
       \rho\bm{u}=\sum_{\alpha=0}^{8}f_{\alpha}\bm{e}_{\alpha}, \quad
       p={c_s^2}\rho,
\end{equation}
where $p$ is the pressure of fluid.

The relaxation time relating to the kinematic viscosity $\nu$ of fluid is given by
\begin{equation}\label{relaxtime}
       \tau=\nu/c_s^2.
\end{equation}
This relationship is $\tau=\nu/c_s^2 + 0.5$ in standard LBM, while the 0.5 is the numerical viscosity comes from the special discretization  scheme to the DBE. The more information can be found in Ref.~\cite{rossi2005unstructured}.

\subsection{Finite volume formulation of LBM}
\begin{figure}
	\centering
	\subfigure{
			\includegraphics[width=0.45 \textwidth]{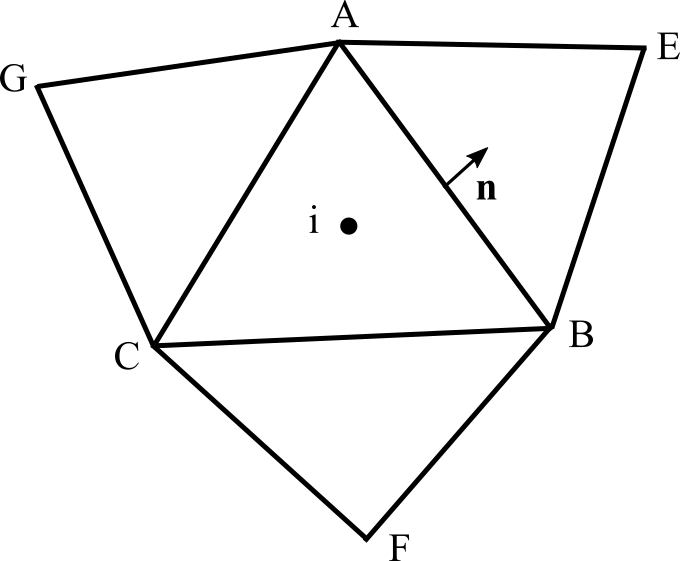}
		}
	\caption{\label{unstructuremesh} The layout of unstructured grid.}
\end{figure}

To discretize the DBE in physical space are the same as traditional FVM. For unstructured grid, the computational domain is discretized into a set of convex polygons, usually triangle, quadrilateral and hybrid of above in 2D. The cell-centered finite volume formulation is used in this paper, that is the distribution function is stored at the center location of polygon. Fig.~\ref{unstructuremesh} is the sample unstructured grid which have four triangle control volumes. The integral form of Eq. \eqref{bgkeq} on the ABC is
\begin{equation}\label{intbgk}
     \frac{\partial}{\partial{t}}{\int_\Omega f_{\alpha}(\bm{x},t) d\Omega} =
     -{\oint_{\partial\Omega} (\bm{e}_\alpha \cdot \bm{n})f_{\alpha,bc} dl}
     -{\int_{\Omega} \frac{1}{\tau}[f_{\alpha}(\bm{x},t) -f_{\alpha}^{eq}(\bm{x},t)] d\Omega},
\end{equation}
where $\Omega$ represent the control volume ABC, $\partial\Omega$ represent the cell interface of ABC, $\bm{n}$ represent the outward-facing unit normal vector of cell face, and $f_{\alpha,bc}$ represent the value of distribution function at the cell interface which need to reconstruct using the distribution function stored at the center of cell.

If ${\bar f}_{\alpha}(\bm{x},t)$ represent the average value of cell, then we have
\begin{equation}
     \int_\Omega f_{\alpha}(\bm{x},t) d\Omega = {\bar f}_{\alpha}(\bm{x},t)A,
\end{equation}
where A is the area of the cell. To use the semi-discrete form, Eq. \eqref{intbgk} can be rewrite as
\begin{equation}\label{semibgk}
      \frac{d}{dt}{\bar f}_\alpha = RHS,
\end{equation}
where $RHS$ represent the Right hand side of Eq. \eqref{intbgk}. The temporal discretization will discuss in the next section. For the spatial discretization, on the control volume, the $RHS$ can be calculated as
\begin{equation}\label{rhs}
      RHS = -\frac{1}{A}\sum_{m}[(\bm{e}_\alpha \cdot \bm{n})f_{\alpha,bc}\Delta{l}]_m
            -\frac{1}{\tau}[f_{\alpha}(\bm{x},t) -f_{\alpha}^{eq}(\bm{x},t)],
\end{equation}
where $m$ is the total number of cell interface and $\Delta{l}$ is the length of the cell interface between cell $i$ and cell $j$.

\begin{figure}
	\centering
	\subfigure{
			\includegraphics[width=0.35 \textwidth]{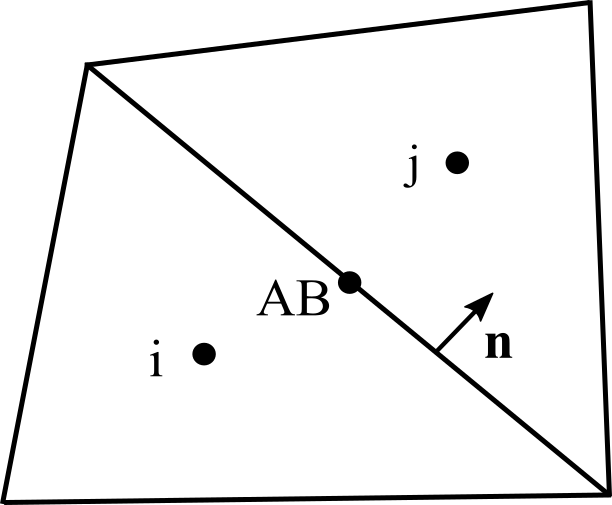}
		}
	\caption{\label{innercell} The reconstruction of distribution functions on the cell interface.}
\end{figure}

In our work, the least squares linear reconstruction (LSLR) upwind scheme developed in Ref.~\cite{stiebler2006upwind} is used to reconstruct the distribution function at cell face. Referring to face $AB$ in Fig \ref{innercell}, the scheme is
\begin{equation}\label{lslrflux}
   f_{\alpha,AB} = \begin{cases}
                    f_{\alpha,i} + \nabla{f_{\alpha,i}} \cdot (\bm{x}_{AB} - \bm{x}_i)  & if \quad \bm{e}_\alpha \cdot \bm{n} > 0, \\
                    f_{\alpha,j} + \nabla{f_{\alpha,j}} \cdot (\bm{x}_{AB} - \bm{x}_j)  & if \quad \bm{e}_\alpha \cdot \bm{n} \leq 0,
                   \end{cases}
\end{equation}
where subscript $AB$, $i$ or $j$ are the centers of the face and the cells, respectively.

To calculate the gradient of distribution function, the least square minimization method is used:
\begin{equation}\label{leastsquare}
      \min_{\nabla{f_{\alpha,i}}}\sum_{n}w_{i,n}[f_{\alpha,n} - f_{\alpha,i} - \nabla{f_{\alpha,i}} \cdot (\bm{x}_n - \bm{x}_i)]^2,
\end{equation}
where $w_{i,n} = 1/(\bm{x}_n - \bm{x}_i)^2$ is the geometrical weighting factor.

\subsection{Time discretization scheme}
To discrete the left hand side of Eq. \eqref{semibgk}, three temporal discretization methods are used in this paper. If ${\bar f}_\alpha^{n+1}$ and ${\bar f}_\alpha^n$ represent the distribution function at time $t+\Delta{t}$ and $t$, respectively, where $\Delta{t}$ is time step, and if we know the ${\bar f}_\alpha^n$, the ${\bar f}_\alpha^{n+1}$ can be calculated by

$(1)$First-order, explicit Euler forward scheme (Euler scheme):
\begin{equation}\label{euler}
    {\bar f}_\alpha^{n+1}= {\bar f}_\alpha^n + \Delta{t}RHS^n,
\end{equation}
where $RHS^n$ is calculated by Eq. \eqref{rhs} with $f_\alpha$ and $f_\alpha^{eq}$ at $n$. As explicit treat the convective term and the collision term, both CFL criterion and stability criterion enforced by collision term will decide the $\Delta{t}$\cite{patil2009finite}. The CFL criterion is\cite{chen2015unified}
\begin{equation}\label{CFL}
   CFL = \frac{\Delta{t}(\left|\bm{e}_\alpha\right|_{max} + \left|\bm{U}\right|_{max})}{\frac{A_{min}}{L_{min}^x + L_{min}^y}},
\end{equation}
where $\left|\bm{U}\right|_{max}$ the maximum macroscopic velocity in the fluid filed, $A_{min}$ is the minimum area of control volume, and $L_{min}^x$ and $L_{min}^y$ are the projected lengths of the minimum control volume in $x$ and $y$ direction, respectively.

The time step decided by collision term is
\begin{equation}\label{timetau}
    \Delta{t} \leq  2\tau.
\end{equation}
The smaller of time step calculated from the Eq. \eqref{CFL} and Eq. \eqref{timetau} is the finial time step.

$(2)$ Explicit Adams-Bashforth scheme (AB2 scheme):
\begin{equation}
      {\bar f}_\alpha^{n+1} = {\bar f}_\alpha^n  + \frac{\Delta{t}}{2}(3RHS^n -RHS^{n-1}).
\end{equation}
The AB2 scheme can achieve second-order temporal accuracy, and Eq. \eqref{euler} can be used in the first iterative step. The largest time step the AB2 scheme can achieve decided by CFL criterion\cite{patil2009finite}.

$(3)$ Explicit four steps Runge-Kutta scheme (RK4 scheme)\cite{deese1988navier}:
\begin{equation}
    \begin{aligned}
        & {\bar f}_\alpha^{(0)} = {\bar f}_\alpha^n, \\
        & {\bar f}_\alpha^{(k+1)} = {\bar f}_\alpha^{(0)} - \beta_{k+1}RHS^{(k)} \quad for \quad k=0,1,2,3,  \\
        & {\bar f}_\alpha^{n+1} = {\bar f}_\alpha^{(4)},
    \end{aligned}
\end{equation}
where $\beta_{1}= 1/4$, $\beta_{2}= 1/3$, $\beta_{3}= 1/2$, and $\beta_{4}= 1$. The RK scheme can much increase the computational efficiency than Euler scheme\cite{ubertini2004recent}.

For these three temporal discretized schemes, the most simple formulation is Euler scheme, but its only first-order accuracy. For AB2, though it can achieve second-order accuracy, it use much memory as it has to store the previous fluxes and time step is about the order of $\tau$. For RK4, it can achieve second-order temporal accuracy at least, and the time step can be greater than that of the Euler scheme, but the computational cost is also larger. So, the advantage of RK4 is it can keep the high-order accuracy at larger time step. In this paper, the Euler scheme is used for steady flow simulations as the temporal accuracy does not impact the results, and the AB2 and RK4 schemes are used for unsteady flow simulations.

\subsection{Boundary condition}\label{Sec:Boundary condition}
How to rational implement the boundary conditions (BCs) are the key issues. Irrational treatment of BCs maybe decline the speed of convergence, even result in the failure of simulation. In this paper, to update the distribution functions $f_\alpha$ stored in the border cells, the ghost cells and special boundary condition treat methods are used and will be illustrated separately later.

\subsubsection{Ghost cells}
For the discretized cells located at the border of computational domain, the number of neighbours of cell will less than the number of interface of cell. For example, if the computational domain is rectangle, the triangular cell maybe have only one neighbour cell at the corner of domain. In this condition, the gradient of $f_\alpha$ calculated from LSLR may have wrong distribution in physical space. To avoid specific treatment of border cell in calculation of the gradient of $f_\alpha$, the idea like ghost cells (GC) are introduced both used to calculate the gradient of $f_\alpha$ and the BCs. For the implement of ghost cell, only one level of ghost cell is defined at the border of computational domain. In general, when ghost cells are used, the flux of boundary face can be calculated same as to the inner face. As the gradient of $f_\alpha$ in the ghost cells are difficult to be reconstructed, the specific scheme are used in this paper.

\begin{figure}
	\centering
	\subfigure{
			\includegraphics[width=0.4 \textwidth]{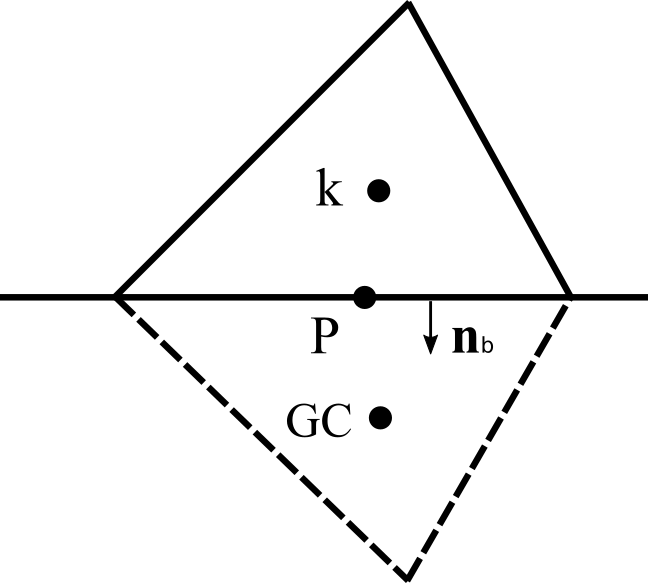}
		}
	\caption{\label{bordercell} The reconstruction of distribution functions on the cell border face.}
\end{figure}

In Fig.~\ref{bordercell}, $k$ is the center of border cell, $P$ is the center of border interface. the $f_\alpha$ in the ghost cells can be calculated as:
\begin{description}
\item[(1)] For the wall boundary, the extrapolation method proposed by Guo et al.\cite{guo2002extrapolation} is used. This scheme has been used in structured and unstructured grid FVLBM\cite{guzel2015simulation,lifinite}. If $P$ represent the center of wall face, the $f_\alpha$ at $P$ can be calculated as

\begin{equation}\label{wallbc}
          f_{\alpha, P} = f_{\alpha}^{eq}(\rho_k,\bm{u}_w) + f_{\alpha, k} - f_{\alpha, k}^{eq},
\end{equation}
where the  $\bm{u}_w$ represents the velocity at wall. For the calculation of the gradient $f_\alpha$ in the cell $i$, the nominal ghost cell isn't needed and this is different to the work of Li et al.'s\cite{lifinite}. As before the time advance to next step, the $f_\alpha$ at $P$ can be calculated from wall boundary condition, it can be used for calculating the gradient of $f_{\alpha,i}$. In the previous works, non-equilibrium bounce-back rule and other schemes developed in standard LBM has been transplant to FVLBM\cite{patil2009finite,peng1999boundary}, the advantage of Guo et al.'s\cite{guo2002extrapolation} compared with other schemes is that it need not to decide which directions are bounce-back to the computational domain and we recommend use this formulation as no-slip wall boundary condition.

\item[(2)] For the inlet boundary, the $f_\alpha$ in the inlet ghost cell can be calculate similar to Eq. \ref{wallbc}:

\begin{equation}
   f_{\alpha, GC} = f_{\alpha}^{eq}(\rho^{*}, u_{\infty}) + f_{\alpha, k} - f_{\alpha, k}^{eq}
\end{equation}
where $\rho^*$ is the density of fluid that needed to be extrapolated, and $u_{\infty}$ is the velocity of free stream. For the external flow around obstacle like cylinder or airfoil, the $\rho^*=\rho_{\infty}$, where $\rho_{\infty}$ is the density of free stream. The reason for this is that it can maintain the mass conservation of computational domain. For internal flow like pipe, the $\rho^*=\rho_i$, that is velocity inlet boundary condition.

\item[(3)] For the outlet ghost cell, if flow is external, the $f_\alpha$ can be straightforward calculate as
\begin{equation}
                  f_{\alpha, GC} = f_{\alpha,i}.
\end{equation}
If flow is internal, it can be calculated as
\begin{equation}
   f_{\alpha, GC} = f_{\alpha}^{eq}(\rho_{\infty}, u_i) + f_{\alpha, k} - f_{\alpha, k}^{eq}
\end{equation}
that is pressure outlet boundary condition as the $p_{\infty}=\rho_{\infty}/3$ in D2Q9 model.

\item[(4)] For symmetric boundary condition, the $f_\alpha$ in ghost cell can be calculated as
\begin{equation}
                  f_{\alpha, GC} = f_{\alpha}^{eq}(\rho_i, u^*,v^*),
\end{equation}
where $u^*,v^*$ are the velocity of fluid that need to be extrapolated. If the symmetric face is located at the horizontal direction, the $u^*, v^*$ can be calculated as $u^*=u_i, v^*=-v_i$. Other directions can be dealt with similar method.
\end{description}

After update the distribution functions in the ghost cell, the gradient of $f_{\alpha,i}$ can be calculated and used for the reconstruction the $f_\alpha$ at the center of inner interface of border cell.

\subsubsection{Boundary treatment}
After update the distribution functions in the ghost cells, the $f$ at the center of border interface will be reconstructed next. For the reconstruction of border interface,
\begin{description}
\item[(1)] For wall interface, the treatment has illustrated in detailed above.

\item[(2)] For inlet interface, the $f$ at $P$ is
\begin{equation}
   f_{\alpha, P} = \begin{cases}
                  f_{\alpha,i} + \nabla f_{\alpha,i} \cdot (\bm{x}_P-\bm{x}_k) & if \quad \bm{e}_\alpha \cdot \bm{n}_b > 0, \\
                  f_{\alpha,GC}  & if \quad \bm{e}_\alpha \cdot \bm{n}_b \leq 0,
                           \end{cases}
\end{equation}
where $\bm{n}_b$ is the outward unit normal vector of border interface.

\item[(3)] For outlet face, for the internal flow, the $f$ at $P$ is
\begin{equation}
   f_{\alpha, P} = \begin{cases}
                  f_{\alpha,i} + \nabla f_{\alpha,i} \cdot (\bm{x}_P-\bm{x}_k)  & if \quad \bm{e}_\alpha \cdot \bm{n}_b > 0, \\
                  f_{\alpha,GC}  & if \quad \bm{e}_\alpha \cdot \bm{n}_b \leq 0,
                           \end{cases}
\end{equation}
for external flow, it can be simply calculated as
\begin{equation}
                  f_{\alpha, P} = f_{\alpha,i},
\end{equation}

\item[(4)] For symmetric interface, the $f$ at $P$ is
\begin{equation}
                  f_{\alpha, P} = \frac{1}{2}(f_{\alpha, i} + f_{\alpha, GC}).
\end{equation}
\end{description}

After the detailed implement method is described above, the general implementation of present FVLBM is presented as follows:
\begin{description}
\item[Step 0.] Initialize $\rho$, $\bm{u}$ on the computational domain, use their values to initialize the $f_{\alpha}^{eq}$ and set $f_{\alpha}=f_{\alpha}^{eq}$ according to Eq. \ref{feq}.

\item[Step 1.] Update the $f$ in the ghost cells and calculate the wall boundary condition given by Eq. \ref{wallbc}.

\item[Step 2.] Compute the gradient of $f_\alpha$ in each cell with Eq. \ref{leastsquare}.

\item[Step 3.] Compute the inlet, outlet and symmetric boundary conditions.

\item[Step 4.] Use Eq. \ref{rhs} to compute the advective and collision terms.

\item[Step 5.] Compute the temporal discretized scheme.

\item[Step 6.] Update the macro qualities $\rho$, $p$ and $\bm{u}$ in each cell.

\item[Step 7.] Go back to Step 1 to start a new iteration.
\end{description}

\subsection{The brief comparisons of present scheme with TVD-FVLBM formulation}
\begin{figure}
	\centering
	\subfigure[]{
			\includegraphics[width=0.34  \textwidth]{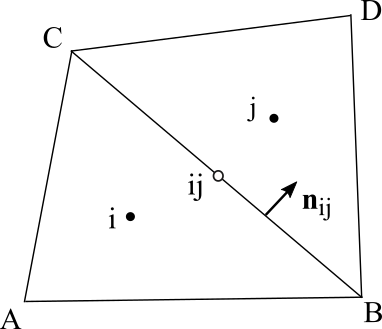}
			}
	\subfigure[]{
			\includegraphics[width=0.45 \textwidth]{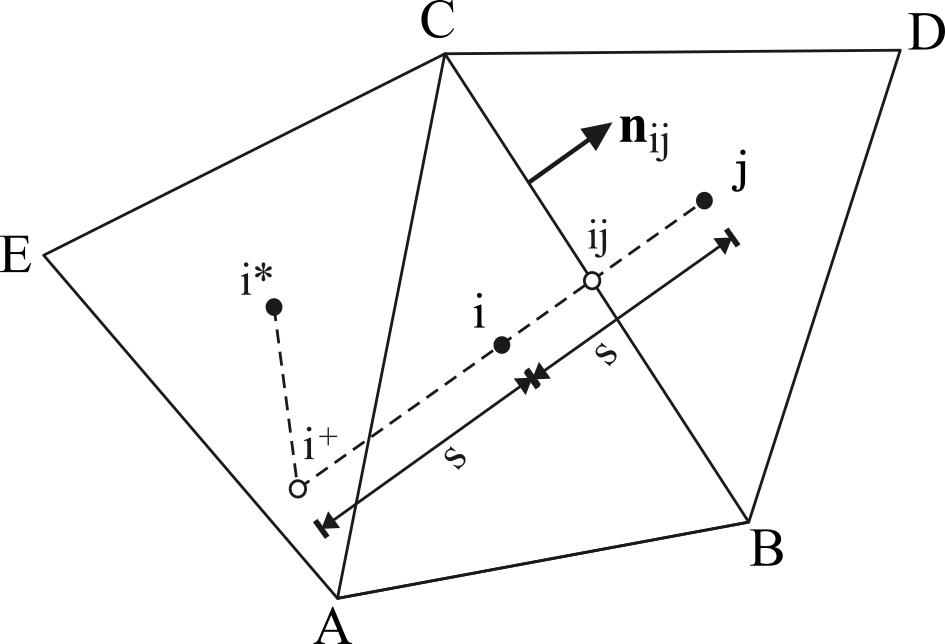}
		}
	\caption{\label{fluxcalstencil} The reconstruction stencil for (a) LSLR scheme and (b) TVD-FVLBM scheme.}
\end{figure}

As TVD-FVLBM scheme presented by Patil et al.\cite{patil2009finite} can be considered as the recent development of FVLBM, here we make a simple comparison between ours and Patil et al.'s. Fig.~\ref{fluxcalstencil} shows the computational stencil for reconstructing the distribution functions $f_\alpha$ at cell interfaces. It is clear that present stencil is compact and it has many advantages for implementing numerical scheme on unstructured grid. On the contrary, the virtual upwind node $i^+$ used in TVD-FVLBM will lead to poor computational efficiency.

In order to make a fair comparison, the explicit Euler scheme given by Eq. \ref{euler} and explicit calculating method for collision terms given by Eq. \ref{rhs} (second term on the right hand side) are used for both schemes, and the gradients of $f_\alpha$ are calculated with Eq. \ref{leastsquare}. The main difference is the reconstruction of $f_\alpha$ at cell interfaces and the brief procedures are described as follow (only the upwind directions based on $\bm{n}_{ij}$ are considered here):

\noindent(a) LSLR scheme:
\begin{description}
\item[Step 1.] Use Eq. \ref{lslrflux} to calculate the $f_\alpha$ at cell interfaces. In order to compare, it will rewrite as
\begin{equation}\label{lslrcompare}
	F(f_{\alpha,ij}) = f_{\alpha,i} + \nabla f_{\alpha,i} \cdot (\bm{x}_{ij}-\bm{x}_i).
\end{equation}
\end{description}

\noindent(b) TVD-FVLBM scheme:
\begin{description}
\item[Step 1.] Calculate the $r$-factor and given by
\begin{equation}\label{rfactor}
       r=-\frac{\bm{d}_{i^*i^+} \cdot \nabla f_{\alpha,i^*}}{f_{\alpha,j} - f_{\alpha,i}} + \frac{f_{\alpha,i} - f_{\alpha,i^*}}{f_{\alpha,j} - f_{\alpha,i}}=\frac{f_{\alpha,i}-\boxed{(f_{\alpha,i^*}+\bm{d}_{i^*i^+} \cdot \nabla f_{\alpha,i^*})}}{f_{\alpha,j} - f_{\alpha,i}},
\end{equation}
where $\bm{d}_{i^*i^+}$ is distance vector and equal to $\bm{d}_{i^*i^+}=\bm{x}^+-\bm{x}^*$.
\item[Step 2.] Calculate the flux limiter and given as (Superbee limiter-function)
\begin{equation}
       \Phi_{ij}(r)=\max\left[0,\min(2r,1),\min(r,2) \right].
\end{equation}
\item[Step 3.] Calculate the left and right states of $f_\alpha$, respectively, and given by
\begin{equation}
   \begin{cases}
   f_\alpha^L=f_{\alpha,i} + \frac{1}{2}\Phi_{ij}(r)(f_{\alpha,j}-f_{\alpha,i}) , \\
   f_\alpha^R=f_{\alpha,j} + \frac{1}{2}\Phi_{ij}(r)(f_{\alpha,i}-f_{\alpha,j}) .
   \end{cases}
\end{equation}
\item[Step 4.] The $f_\alpha$ at cell interfaces can be calculated as
\begin{equation}
       F(f_{\alpha,ij})=\frac{1}{2}\left[F(f_\alpha^R)+F(f_\alpha^L)-\left|\bm{e}_\alpha \cdot \bm{n}_{ij} \right|(f_\alpha^R-f_\alpha^L)\right].
\end{equation}
\end{description}

It is clear that the computational costs of terms in box of Eq. \ref{rfactor} are equal to that of LSLR scheme (Eq. \ref{lslrcompare}), and all of other terms used in TVD-FVLBM need additional computational costs. Besides, further additional computational costs are needed to improve the performance of TVD-FVLBM scheme for simulating the unsteady flows\cite{patil2012two}. Next section we will prove that the present simple scheme is good enough for simulating the basic steady and unsteady incompressible laminar flows, the original and improved TVD-FVLBM scheme may apply to special incompressible or compressible flows.

\section{Numerical experiments}\label{Sec:Numerical experiments}
In this section, four flow problems are simulated to validate the method used in this paper. One plate driven Couette flow which have analytical solution is chosen as the first case to test the temporal and spatial accuracy. Then, three laminar flow cases, namely lid-driven square cavity flow, flow around single circular cylinder, and around more complex double circular cylinders (tandem and side-by-side with different spacing), are solved to validate the code developed in this paper. The unstructured grid used in this section are generated through Salome, an open-source software that provides a generic platform for pre- and post-processing for numerical simulation (\url{http://www.salome-platform.org/}).

Before the discussions of numerical results, the framework of code is outlined briefly. FVLBM formulation has been coded with the help of Code\underline{ }Saturne, an open-source computational fluid dynamics (CFD) software of Electricite De France (EDF), France (\url{http://code-saturne.org/cms/}). Code\underline{ }Saturne solves the incompressible NS equations, and coupling with several turbulence models for turbulent flow simulation\cite{archambeau2004code}. In our work, the kernel code of the present FVLBM is developed in Fortran language and incorporated into the software, where other codes, such as the mesh data structure, is still followed the framework of Code\underline{ }Saturne, and a switch is added to choose the FVLBM solver or Navier-Stokes solver. The parallel computational modules in the software are also used to make the code of FVLBM have the ability of parallel computing.

\subsection{One plate driven Couette flow}
\begin{figure}
	\centering
	\subfigure{
			\includegraphics[width=0.45 \textwidth]{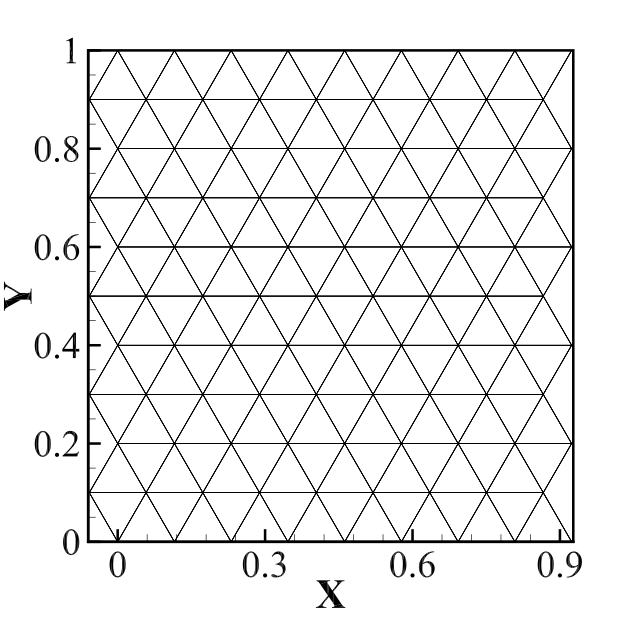}
		}
	\caption{\label{couettemesh} The mesh used for one plate driven Couette flow which have $10$ cells in $y$-direction.}
\end{figure}

To evaluate the temporal and spatial accuracy of FVLBM, the simulation of the plate Couette flow is carried out. In this problem, the top plate is set to move with constant velocity $U$ along $x$-direction, and the bottom plate is stationary. The wall boundary condition is applied to two plates, and the periodic boundary condition is applied in $x$-direction. Fig.~\ref{couettemesh} shows one of the grids used in this case. The computational domain is discretized with equilateral triangle. 16 triangles are placed in $x$-direction and a set of triangles, from 5 triangles to 80 triangles are placed in $y$-direction. The Mach number Ma is 0.1 and $U=c_{s}$Ma. The fluid density $\rho$ is 1.0 and the Reynolds number $Re = UL/\nu$ is 10, where $L$ is the channel width. For unsteady Couette flow, the expression of analytical solution can be found in Ref.~\cite{lee2001characteristic}, where the velocity profile is change depended on the time. For steady Couette flow, where the velocity profile depended on $U$ and $L$ ($u=yU/L$) is used to test spatial accuracy.

\begin{figure}
	\centering
	\subfigure{
	        \label{timeerror}
			\includegraphics[width=0.45 \textwidth]{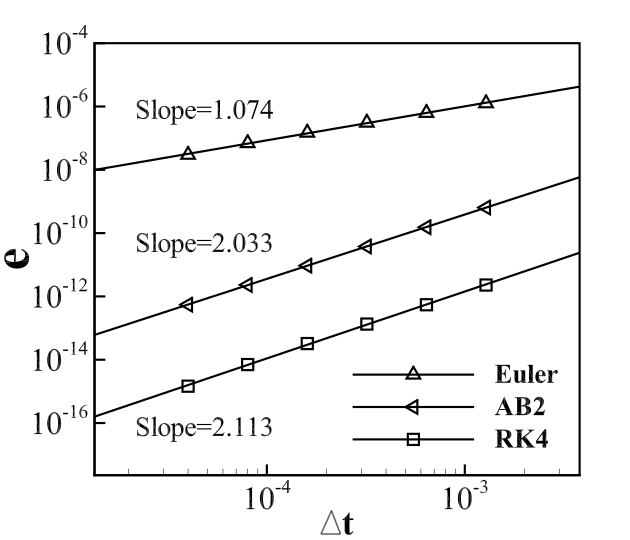}
		}
	\subfigure{
	        \label{spatialerror}
			\includegraphics[width=0.45 \textwidth]{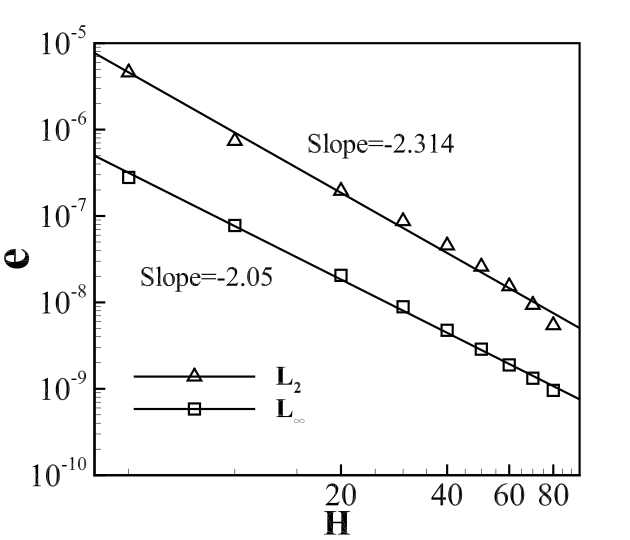}
		}
	\caption{\label{fvlbmerror} The (a) $L_\infty$ norms of temporal error and (b) $L_2$ and $L_{\infty}$ norms of spatial error for the simulation of plate Couette flow.}
\end{figure}

To eliminate the impact of spatial error on the temporal error, the temporal error is computed by comparing the solutions of different temporal resolutions to a referenced solution at the same spatial resolution (80 triangles in $y$-direction), where the referenced solution is calculated at a very small time step ($\Delta t=2\times10^{-5}$). The $L_\infty$ norms of temporal error $e$ is computed at $t=8$ and at the center vertical line of mesh. From Fig.~\ref{timeerror} ($L_\infty$ norms) we can find, the first Euler forward scheme is first-order accuracy, the AB2 and RK4 both can achieve second-order accuracy. Beside, for RK4, we also test the temporal error at different advancing coefficients, the only first-order accuracy can achieved at other coefficients. Due to the limited knowledge of the authors, maybe the accuracy of spatial discretization is only second-order in theory, the RK4 can't achieve fourth-order accuracy in temporal discretization. The advantage of use RK4 as temporal discretized method is that to achieve same accuracy, the computational efficiency is much large than Euler scheme\cite{ubertini2004recent}. Besides, the largest time step that the AB2 can achieve maybe restrict both CFL condition and the numerical stability of collision term. That is contradict with Ref.~\cite{patil2009finite} which is only decided by CFL condition. From our numerical test, the largest time step maybe $\Delta{t}\leq\tau$. The detailed numerical analysis same as to Ref.~\cite{misztal2015detailed} is needed to solve this puzzle.

For the analysis of spatial error, at the same small time step $\Delta{t}=2\times10^{-4}$, the resolution of grid $H$ at $y$-direction is changed continuous to compare the error $e$ between the numerical result and the analytic solution. Fig.~\ref{spatialerror} shows the $L_2$ and $L_\infty$ norms of spatial error calculated at the center line of computational domain in $x$-direction and at time $t=8.0$. It can be found that both $L_2$ norms and $L_\infty$ norms show the spatial error are around second-order. That means the method of LSLR upwind scheme can achieve second-order accuracy.

\begin{figure}
	\centering
	\subfigure{
			\includegraphics[width=0.5 \textwidth]{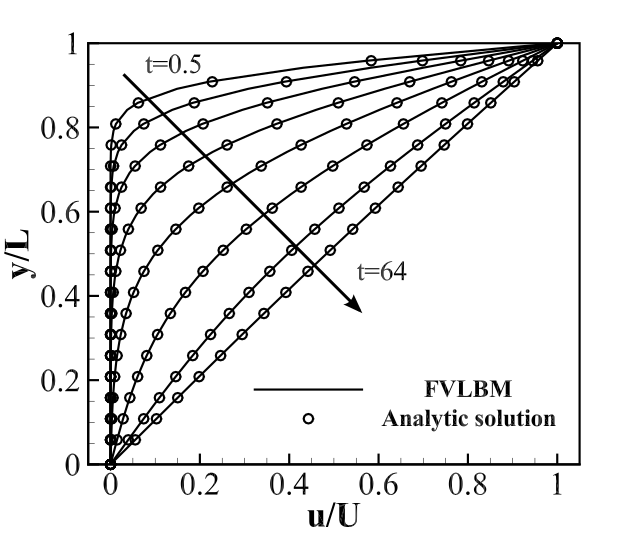}
		}
	\caption{\label{couettevelocity} Comparison of normalized velocity profiles with analytical solution in the center line of channel at $x$-direction with different times.}
\end{figure}

Finally, we compare the numerical results with the analytic solutions at different time $t$, which the time step is $\Delta{t}=2\times10^{-4}$ and two grids are chose which the resolution at $y$-direction are 20 (same as to Lee et al.\cite{lee2001characteristic}) and 40 triangles, respectively, AB2 is used as the temporal discretized scheme. Fig.~\ref{couettevelocity} shows the comparison of normalized velocity profile with 40 triangles at $y$-direction, demonstrating good consistency. The maximal error between analytic solutions and numerical results at different time are decline as time goes on and at $t=0.5$ are about $1.94\times10^{-3}$ (for $20$ triangles) and $6.65\times10^{-4}$ (for $40$ triangles), respectively. The same case also can be found in Ref.~\cite{patil2009finite}, which set $256$ triangles in $y$-direction and the error at different times are about $0.4-0.7\%$. It means that to achieve same accuracy, the grid used in this paper is little compared with Patil et al.'s.

\subsection{Lid-driven square cavity flow}
In this case, we investigate the lid-driven square cavity flow. Depending on the Reynolds number, a set of vortexes with different scales can be found in cavity. As a benchmark problem, both macro method\cite{ghia1982high} and LBM\cite{hou1994simulation, ubertini2004recent} have been studied extensively. In cavity, all borders are defined as wall boundary condition, one move with constant velocity $U$ and others are stationary. The Reynolds number is defined as $Re = UL/\nu$, where $L$ is the cavity width, and $\nu$ is the kinematic viscosity of fluid in the cavity. Four different Reynolds numbers ranging from $400$ to $5000$ are simulated ($Re = 400, 1000, 3200$ have been studied in Ref.~\cite{patil2009finite}, more higher $Re = 5000$ is also study in this paper). Since in this range of Reynolds number, the flow is laminar and steady, the Euler scheme is used as temporal discretization to decline the computational cost.

\subsubsection{Grid-convergence and computational efficiency studies}
\begin{figure}
	\centering
	\subfigure[]{
			\label{mesh32:a}
			\includegraphics[width=0.45 \textwidth]{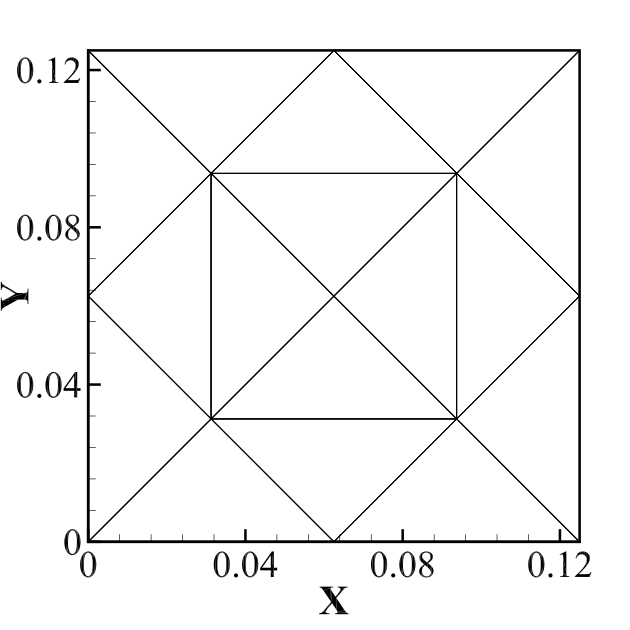}
		}
    \subfigure[]{
    		\label{mesh32:b}
    		\includegraphics[width=0.45 \textwidth]{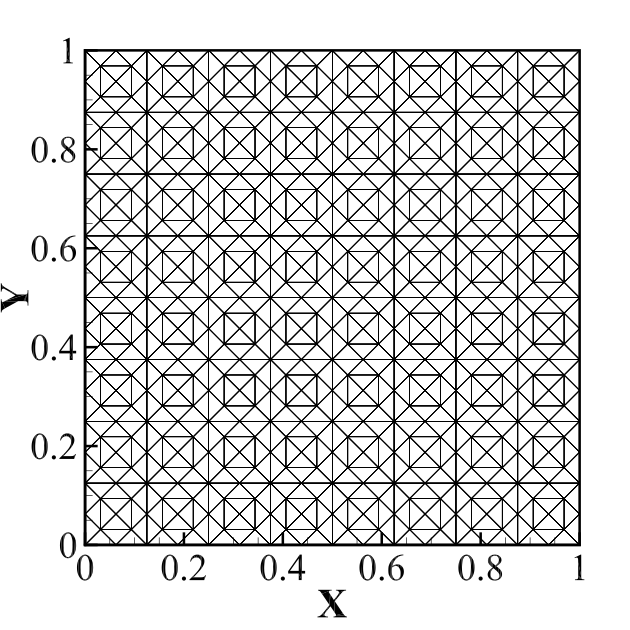}
    	}
	\caption{\label{mesh32} Mesh $32^2$ for grid convergence study: (a) One discretized square cell, (b) Full domain.}
\end{figure}

\begin{figure}
	\centering
	\subfigure[]{
			\label{cavityconvergence-u}
			\includegraphics[width=0.45 \textwidth]{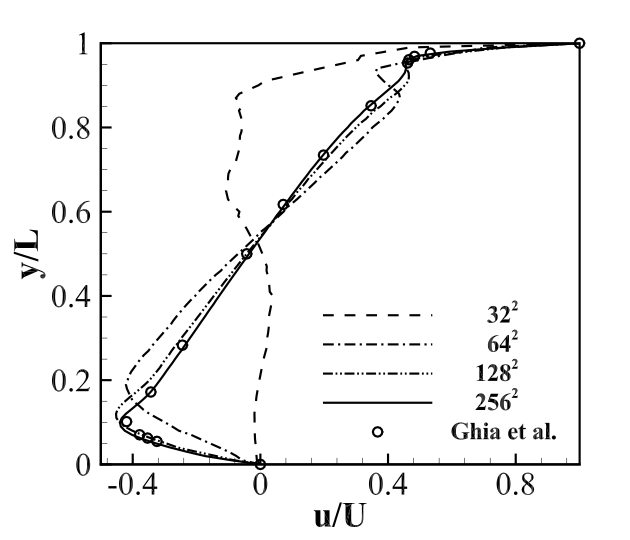}
		}
    \subfigure[]{
			\label{cavityconvergence-v}
			\includegraphics[width=0.45 \textwidth]{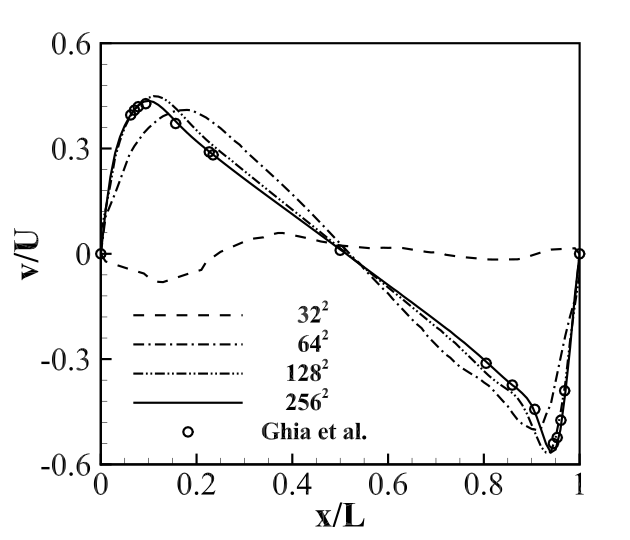}
		}
	\caption{\label{meshconvergencestudy} The comparison of velocity profile between FVLBM and Ghia et al.\cite{ghia1982high} with the location extracted from (a) the horizontal and (b) the vertical centerline of cavity in grid convergence study with $Re = 3200$.}
\end{figure}

\begin{figure}
	\centering
	\subfigure[]{
			\label{convergence-uRe3200}
			\includegraphics[width=0.45 \textwidth]{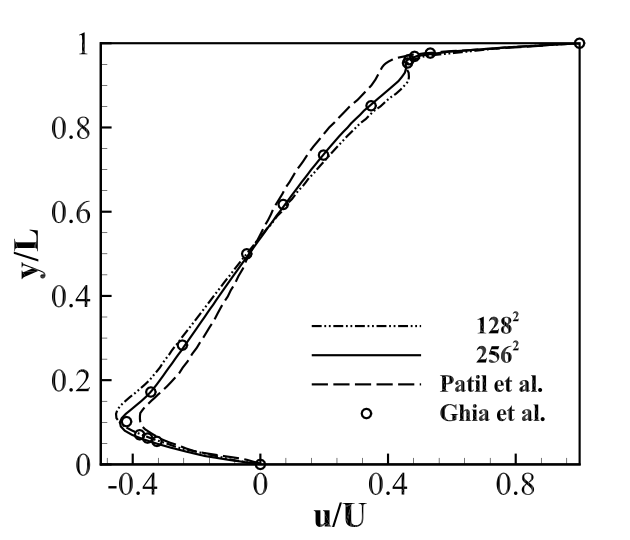}
		}
    \subfigure[]{
			\label{convergence-vRe3200}
			\includegraphics[width=0.45 \textwidth]{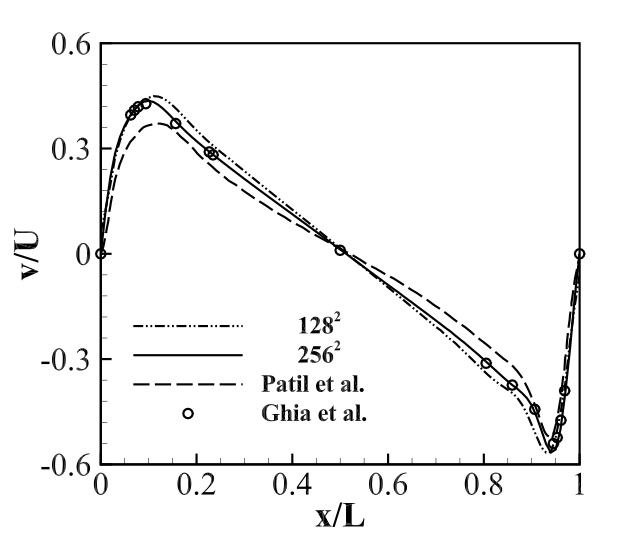}
		}
	\caption{\label{convergence-Re3200} The comparison of velocity profile between FVLBM, Patil et al.\cite{patil2009finite} and Ghia et al.\cite{ghia1982high} with the location extracted from (a) the horizontal and (b) the vertical centerline of cavity in grid convergence study with $Re = 3200$.}
\end{figure}

\begin{figure}
	\centering
	\subfigure{
			\includegraphics[width=0.5 \textwidth]{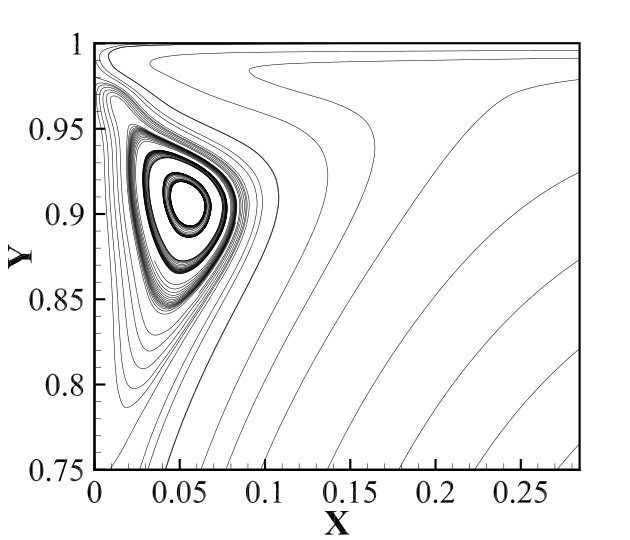}
		}
	\caption{\label{convergence-streamlinesRe3200} The streamlines of cavity flow at top-left corner of geometry for $Re = 3200$.}
\end{figure}

First, the grid convergence studies have been carried out by simulating the flow at $Re = 3200$ with the same mesh system presented in Ref.~\cite{patil2009finite}. Fig.~\ref{mesh32} shows one of the grid system, the cavity is discretized into $8^2$, $16^2$, $32^2$ and $64^2$ squares, respectively, and one square is discretized into 16 right triangles, so the total number of triangles are $32^2$, $64^2$, $128^2$ and $256^2$, respectively. The lid velocity is set to $u=0.1$, $v=0$, and remaining computational domain has been initialized with $u=0$, $v=0$. The time step is $\Delta{t}=3\tau/2$. Fig.~\ref{meshconvergencestudy} shows the normalized $u$ and $v$ velocity profile at the centerline of cavity in $x$- and $y$-direction. With the mesh refined successive, the results can achieve the expected accuracy corresponding to Ghia et al.'s\cite{ghia1982high}. Fig.~\ref{convergence-Re3200} shows the comparison of velocity profile between our and Patil et al.'s\cite{patil2009finite}, where the results of Patil are extracted from the most finest grid $256^2$. It can be found, the second finest grid used in this case have much better results than Patil et al.'s finest grid. The explain from Patil et al. is to calculate the gradient of the $f$ with first-order will introduce much error. But, from this paper we prove that the first-order scheme is adequate accuracy for calculation the gradient of $f$. So, maybe the wall boundary condition and other reasons decline the accuracy of TVD-FVLBM presented by Patil et al.. For the most finest grid $256^2$, although the FVLBM can obtain more accuracy results of velocity profile, from Fig.~\ref{convergence-streamlinesRe3200} we can find that the structure of left and top secondary eddy is a little strange, the margin of eddy isn't smooth compared with the result of Ghia et al.'s\cite{ghia1982high}, that maybe the scheme presented in this paper will decline the accuracy when the right triangles are used. In Table \ref{tab:cavitygridcon}, we also compare the location of centers of primary eddy and corner eddies. It is evident that we obatained more accuracy result than Patil et al.'s, especially for the left top eddy.

\begin{table}\tiny
	\centering
	\caption{\label{tab:cavitygridcon} The location of the centers of primary eddy and corner eddies at $Re = 3200$.}
	\begin{tabular}{ccccccccc}
       \toprule
		 \multicolumn{1}{c}{\multirow{2}{*}{Work}}& \multicolumn{2}{c}{Primary} & \multicolumn{2}{c}{Lower left} & \multicolumn{2}{c}{Lower right} &  \multicolumn{2}{c}{Upper left} \\
		 \cline{2-3}
		 \cline{4-5}
		 \cline{6-7}
		 \cline{8-9}
		  ~& \multicolumn{1}{c}{$x$} & \multicolumn{1}{c}{$y$} & \multicolumn{1}{c}{$x$} & \multicolumn{1}{c}{$y$} &
		     \multicolumn{1}{c}{$x$} & \multicolumn{1}{c}{$y$} & \multicolumn{1}{c}{$x$} & \multicolumn{1}{c}{$y$} \\
	   \midrule
		Ghia et al.\cite{ghia1982high} & 0.5165 & 0.5469 & 0.0859 & 0.1094 & 0.8125 & 0.0859 & 0.0547 & 0.8984 \\
		Patil et al.\cite{patil2009finite} & 0.5189 & 0.5441 & 0.0993 & 0.0963 & 0.8619 & 0.0971 & 0.0316 & 0.8689 \\
		Present, mesh $128^2$ & 0.5186 & 0.5399 & 0.0868 & 0.1138 & 0.8356 & 0.0917 & 0.0458 & 0.9353 \\
		Present, mesh $256^2$ & 0.5157 & 0.5386 & 0.0813 & 0.1178 & 0.8257 & 0.0860 & 0.0549 & 0.9069 \\
       \bottomrule
	\end{tabular}
\end{table}

\begin{figure}
	\centering
	\subfigure[]{
	        \label{fvlbmtvdpresent}
			\includegraphics[width=0.45 \textwidth]{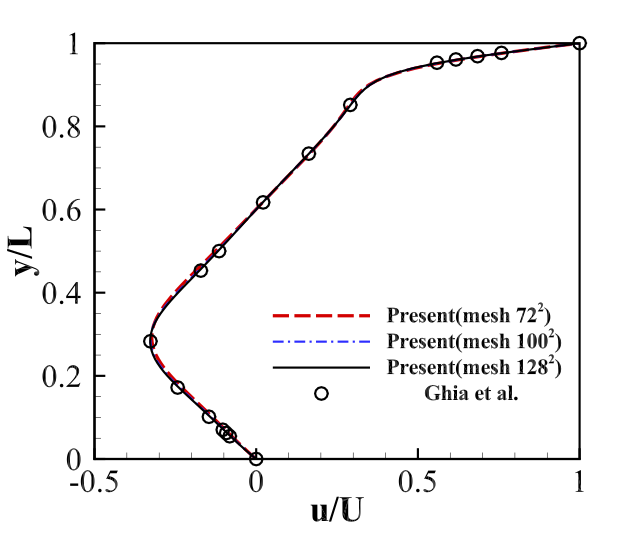}
		}
	\subfigure[]{
	        \label{fvlbmtvdminmod}
			\includegraphics[width=0.45 \textwidth]{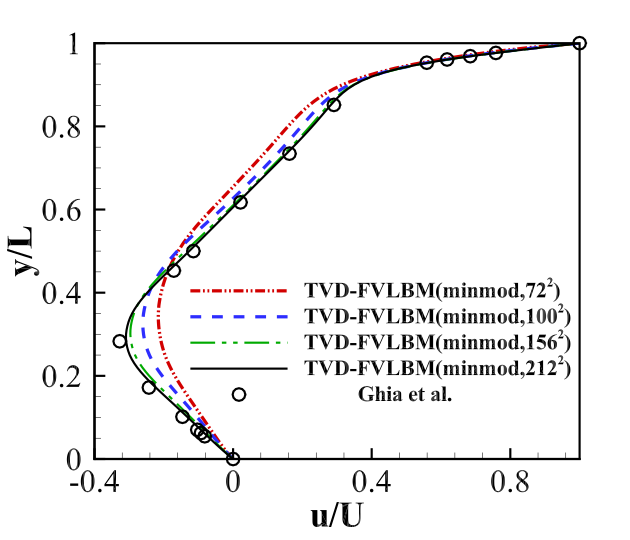}
		}
	\subfigure[]{
	        \label{fvlbmtvdsuperbee}
			\includegraphics[width=0.45 \textwidth]{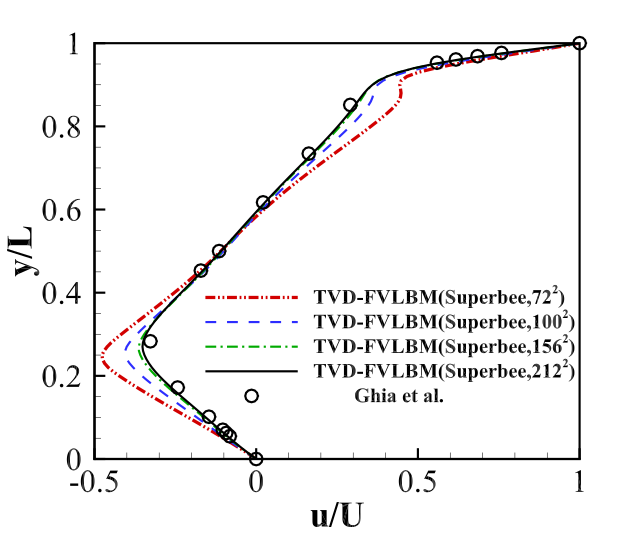}
		}
	\subfigure[]{
         	\label{fvlbmtvdcompare}
			\includegraphics[width=0.45 \textwidth]{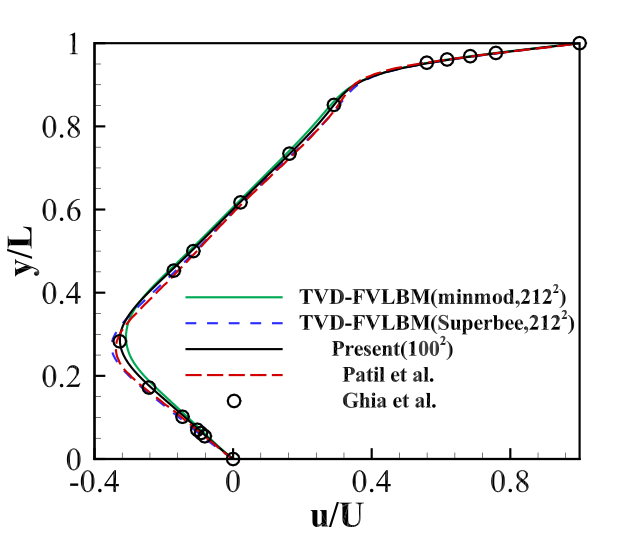}
		}
	\caption{\label{fvlbmtvdRe400} The comparison of $u$ velocity profiles with the location at the horizontal centerline of cavity and $Re = 400$. The data are obtained by (a) present scheme, (b)(c), Patil et al.'s\cite{patil2009finite} TVD-FVLBM scheme with minmod and Superbee limiters, respectively, and (d) Patil et al.'s original data presented in Ref.~\cite{patil2009finite}.}
\end{figure}

\begin{figure}
	\centering
	\subfigure{
			\includegraphics[width=0.5 \textwidth]{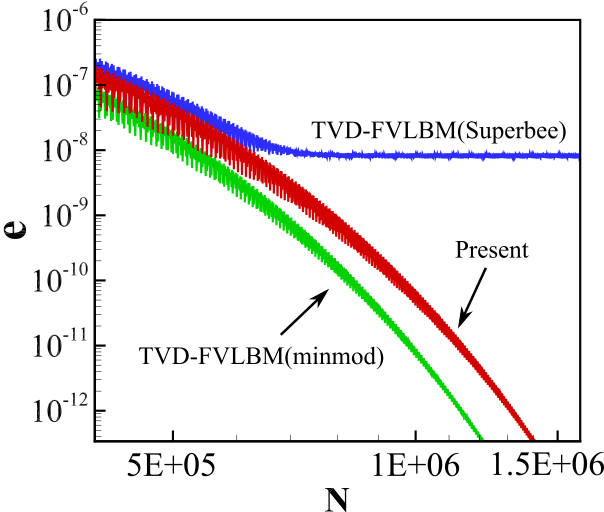}
		}
	\caption{\label{fvlbmtvdresidual} The convergence history of velocity residual $e$ at every 1000 iteration $N$ for lid-driven square cavity flow with grid cells$100^2$.}
\end{figure}

In order to further compare the computational efficiency between the present scheme and Patil et al.'s TVD-FVLBM scheme\cite{patil2009finite}, the TVD-FVLBM scheme is reproduced in our work. To reflect the characteristics of TVD scheme and eliminate the influence of grid, the non-uniform Cartesian grid is used to perform the simulations again at Reynolds number $Re = 400$ (the distance vectors $\bm{d}_{i^*i^+}$ in Eq. \ref{rfactor} are equal to zero if uniform Cartesian grid is used). In this subsection, six different grids are generated with the number of quadrangular elements are equal to $72^2$, $100^2$, $128^2$, $156^2$, $184^2$ and $212^2$, respectively. Fig.~\ref{fvlbmtvdpresent} shows the evolution of $u$ velocity profiles with grid refinement. It is clear that about 10 thousand elements are enough to resolve the flow features with present scheme. On the contrary, the TVD-FVLBM scheme need about 40 thousand elements to achieve the same effects (see Figs.~\ref{fvlbmtvdminmod}-\ref{fvlbmtvdsuperbee}). Fig.~\ref{fvlbmtvdcompare} shows the comparison of different schemes at their greatest results. Generally speaking, the TVD scheme with Superbee limier-function can obtain more good results than that of minmod limier-function. With the same Superbee limiter-function, the results obtained from $212^2$ number of non-uniform grid cells are very close to Patil et al.'s original data with 56528 number of triangular elements. Though much larger number of grid cells are used, the results obtained from TVD-FVLBM scheme are not agree well with Ghia et al.'s then the results obtained from scheme used in this paper. Besides, the present scheme also shows good convergence property of velocity residual $e$ (see Fig \ref{fvlbmtvdresidual}). The $e$ can be given by

\begin{equation}
       e=\frac{\sqrt{\sum_{i}\left[({u_i^{n+1000}-u_i^n})^2+({v_i^{n+1000}-v_i^n})^2\right]}}{\sqrt{\sum_{i}(u_i^n+v_i^n)}},
\end{equation}
where $i$ is the index number of cells. In contrast, the TVD-FVLBM scheme with Superbee limiter-function shows poor convergence property, and similar phenomenons also be found by Patil et al.\cite{patil2009finite}. Finally, Table \ref{tab:cavitycaltime} lists the computational times for different schemes, the total iteration number is $10^5$, and the number of grid cells is $72^2$. The data are obtained from serial computational simulations on a workstation with two Intel(R) Xeon(R) CPUs(2.60GHz, six-core). Finally, compared with TVD-FVLBM formulation, the present scheme only use about $1/4$ number of grid cells and about $80\%$ computational costs for each iteration, that is only use $20\%$ computational amount can obtain more accuracy results in this test case.

\begin{table}
    \centering
	\caption{\label{tab:cavitycaltime} The CPU time costs for $10^5$ iteration steps using different numerical schemes to simulate lid-driven square cavity flow with $Re = 400$ and $72^2$ number of grid cells.}
	\begin{tabular}{cc}
	  \toprule
	    \multicolumn{1}{c}{Numerical scheme} & \multicolumn{1}{c}{CPU costs(sec)}  \\
	  \midrule
	     Present               & 576.60     \\
	     TVD-FVLBM(minmod)     & 697.60     \\
	     TVD-FVLBM(superbee)   & 738.76     \\
	  \bottomrule
	\end{tabular}
\end{table}

\subsubsection{Flow simulations at different Reynolds number with general grid}
\begin{figure}
	\centering
	\subfigure{
			\includegraphics[width=0.5 \textwidth]{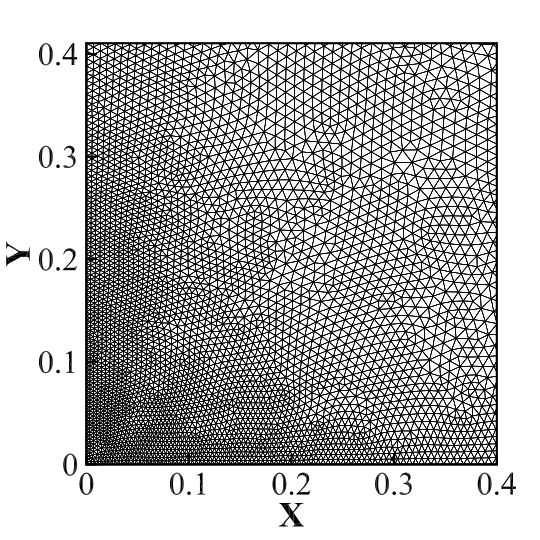}
		}
	\caption{\label{cavitymeshallRe} The part of grid used for lid-driving cavity flow.}
\end{figure}

\begin{figure}
	\centering
	\subfigure[]{
			\label{cavitystreamlinesRe400}
			\includegraphics[width=0.45 \textwidth]{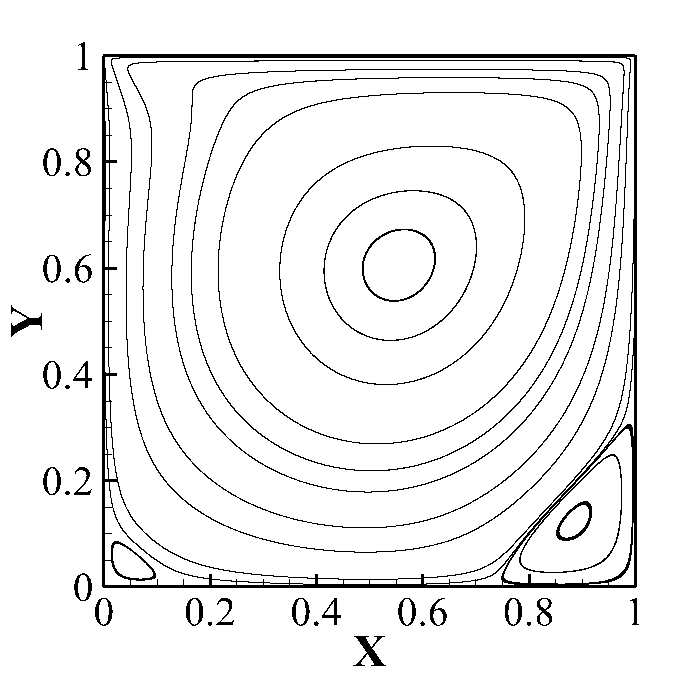}
		}
    \subfigure[]{
			\label{cavitystreamlinesRe1000}
			\includegraphics[width=0.45 \textwidth]{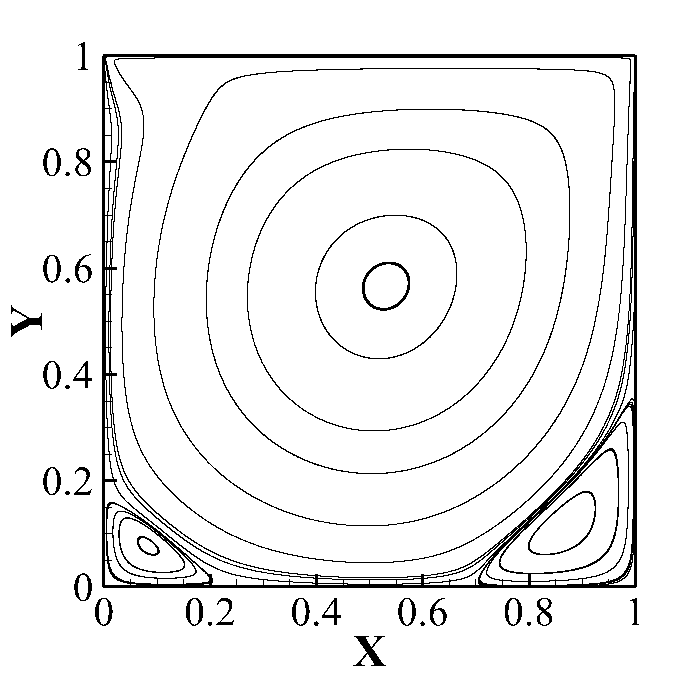}
		}
	\subfigure[]{
			\label{cavitystreamlinesRe3200}
			\includegraphics[width=0.45 \textwidth]{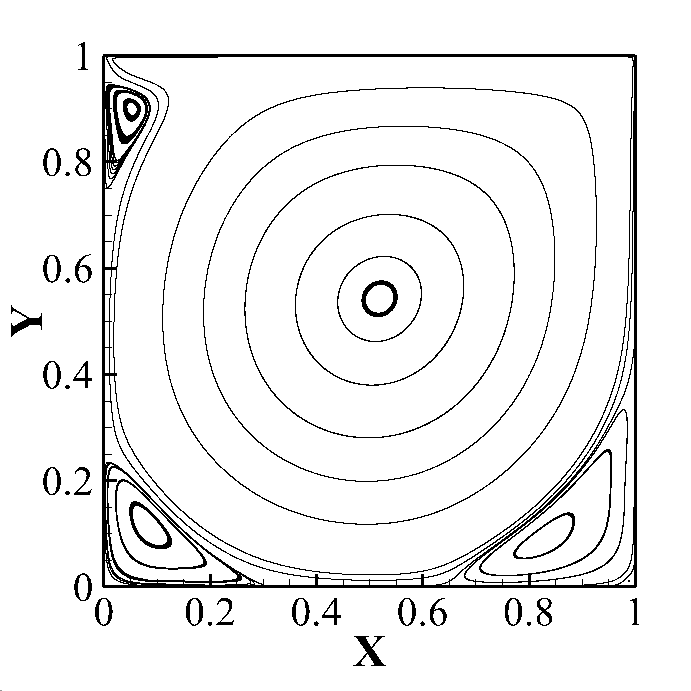}
		}
    \subfigure[]{
			\label{cavitystreamlinesRe5000}
			\includegraphics[width=0.45 \textwidth]{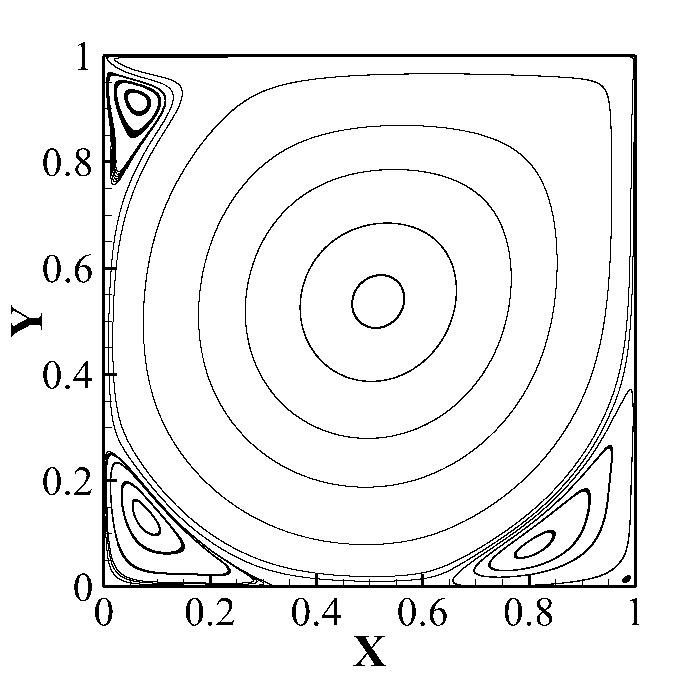}
		}
	\caption{\label{cavitystreamlines} The streamlines of lid-driving cavity flow for (a) $Re = 400$, (b) $Re = 1000$, (c) $Re = 3200$ and (d) $Re =5000$.}
\end{figure}

\begin{figure}
	\centering
	\subfigure[]{
			\label{cavitysRe400-u}
			\includegraphics[width=0.3 \textwidth]{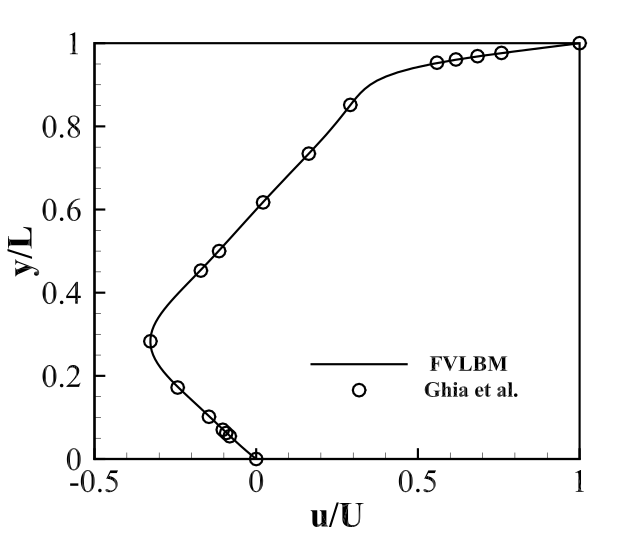}
			\label{cavitysRe400-v}
			\includegraphics[width=0.3 \textwidth]{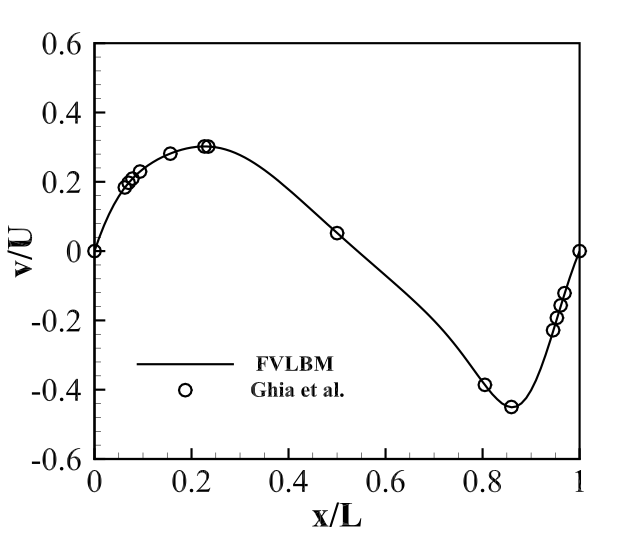}
		}
    \subfigure[]{
			\label{cavitysRe1000-u}
			\includegraphics[width=0.3 \textwidth]{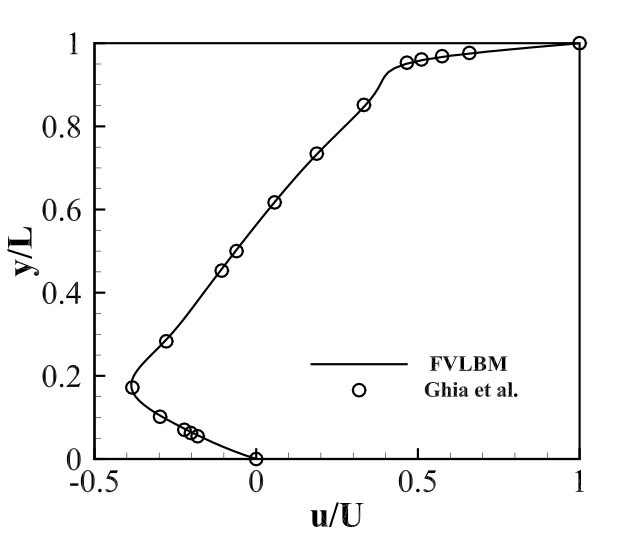}
			\label{cavitysRe1000-v}
			\includegraphics[width=0.3 \textwidth]{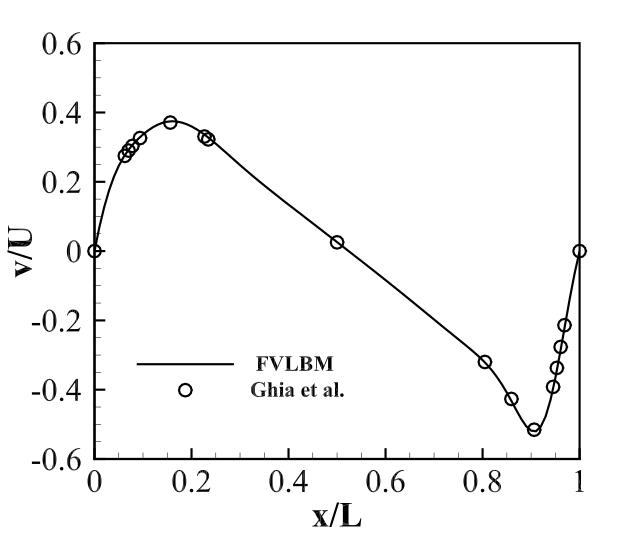}
		}
    \subfigure[]{
			\label{cavitysRe3200-u}
			\includegraphics[width=0.3 \textwidth]{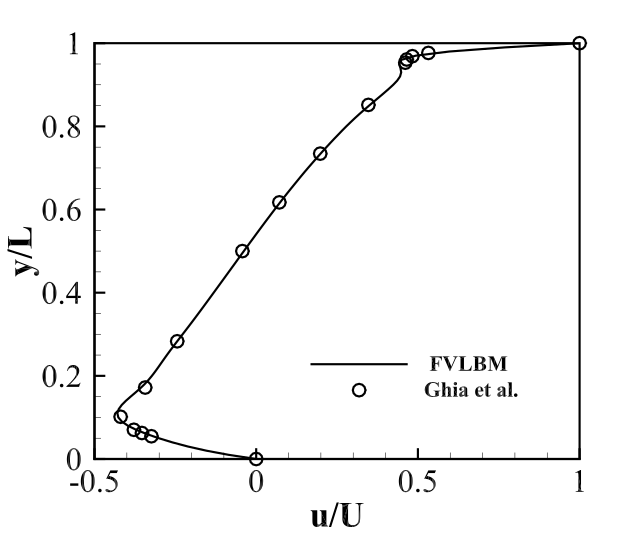}
			\label{cavitysRe3200-v}
			\includegraphics[width=0.3 \textwidth]{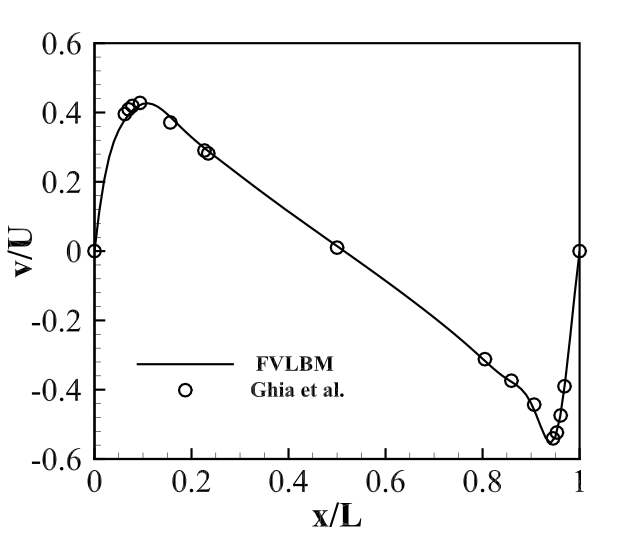}
		}
    \subfigure[]{
			\label{cavitysRe5000-u}
			\includegraphics[width=0.3 \textwidth]{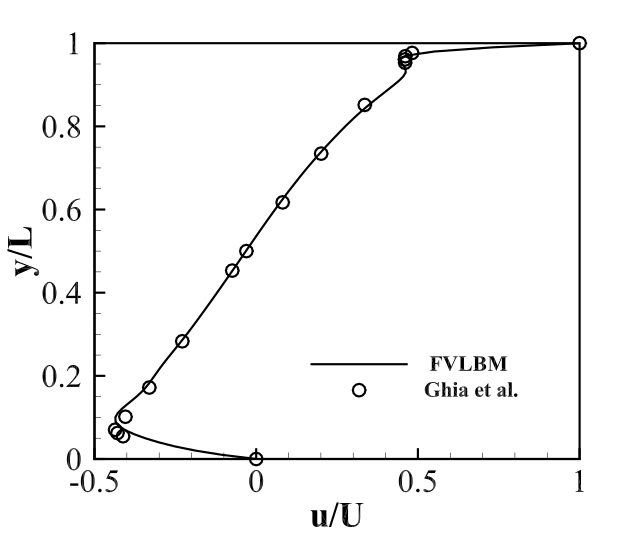}
			\label{cavitysRe5000-v}
			\includegraphics[width=0.3 \textwidth]{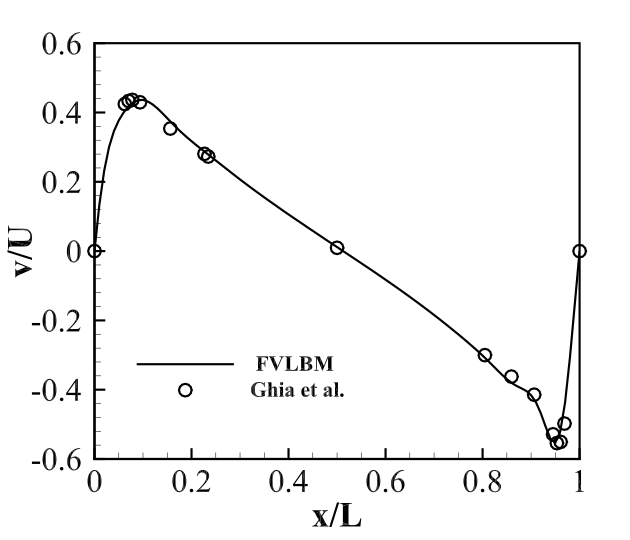}
		}
	\caption{\label{cavityuvallRe} The comparison of velocity profile between FVLBM and Ghia et al.\cite{ghia1982high} with the location extracted from the horizontal and the vertical centerline of cavity for (a) $Re = 400$, (b) $Re = 1000$, (c) $Re = 3200$ and (d) $Re = 5000$.}
\end{figure}

Next, we use general mesh to simulate the cavity flow at $Re = 400$, $1000$, $3200$ and $5000$. The strategy to construct the mesh is all control volume are acute triangle, and refined at the corners of square. Through several tests, we find that the total number of triangle around $128^2$ which used in Ghia et al.\cite{ghia1982high} for $Re\leq3200$ can not achieve expected accuracy. For top left eddy at $Re = 3200$, if the mesh is too coarse, both the scale and the shape of eddy will have strange result. Besides, for $Re = 3200$, the least size of cells around the top-left corner less than $0.004L$ will have good results. Fig.~\ref{cavitymeshallRe} shows the part of grid used in this case finally, the number of triangles is $36502$ (the amount in Ref.~\cite{patil2009finite} is 56528). Fig.~\ref{cavitystreamlines} presents the streamlines in cavity at four different Reynolds numbers studied in this case. Fig.~\ref{cavityuvallRe} shows the normalized velocity profile at the centerline of cavity at different Reynolds numbers, we have good results corresponding to Ghia et al.'s\cite{ghia1982high} at $Re \leq {3200}$. But for $Re = 5000$, the present grid seems too coarse to capture the accuracy results. Table \ref{tab:cavityRe} lists the location of the center of the primary eddy and the corner eddy. It is clear that the formulation used in this paper can capture the correct location of eddy at different Reynolds numbers. From Table \ref{tab:cavityeddyscales}, we can obtain the good results for Re$\leq3200$, but for $Re = 5000$, the size of vortexes are less the results of Ghia et al.'s.

\begin{table}\tiny
	\centering
	\caption{\label{tab:cavityRe} The location of the centers of primary eddy and corner eddies at different Reynolds numbers.}
	\begin{threeparttable}
	\begin{tabular}{cccccccccc}
       \toprule
		 \multicolumn{1}{c}{\multirow{2}{*}{Re}} & \multicolumn{1}{c}{\multirow{2}{*}{Reference}} & \multicolumn{2}{c}{Primary} &      \multicolumn{2}{c}{Lower left} & \multicolumn{2}{c}{Lower right} &  \multicolumn{2}{c}{Upper left} \\
		 \cline{3-4}
		 \cline{5-6}
		 \cline{7-8}
		 \cline{9-10}
		  ~&~& \multicolumn{1}{c}{$x$} & \multicolumn{1}{c}{$y$} & \multicolumn{1}{c}{$x$} & \multicolumn{1}{c}{$y$} &
		       \multicolumn{1}{c}{$x$} & \multicolumn{1}{c}{$y$} & \multicolumn{1}{c}{$x$} & \multicolumn{1}{c}{$y$} \\
	   \midrule
		\multicolumn{1}{c}{\multirow{4}{*}{400}} & a & 0.5547 & 0.6055 & 0.0508 & 0.0469 & 0.8906 & 0.1250 &
        	\multicolumn{1}{c}{---} & \multicolumn{1}{c}{---} \\
		~& b & 0.5608 & 0.6078 & 0.0549 & 0.0510 & 0.8902 & 0.1255 & \multicolumn{1}{c}{---} & \multicolumn{1}{c}{---} \\
		~& c & 0.5506 & 0.5972 & 0.0526 & 0.0471 & 0.8862 & 0.1258 & \multicolumn{1}{c}{---} & \multicolumn{1}{c}{---} \\
		~& d & 0.5543 & 0.6061 & 0.0507 & 0.0471 & 0.8859 & 0.1225 & \multicolumn{1}{c}{---} & \multicolumn{1}{c}{---} \\
		
		\multicolumn{1}{c}{\multirow{4}{*}{1000}} & a & 0.5313 & 0.5625 & 0.0859 & 0.0781 & 0.8594 & 0.1094 &
		     \multicolumn{1}{c}{---} & \multicolumn{1}{c}{---} \\
		~& b & 0.5333 & 0.5647 & 0.0902 & 0.1059 & 0.8667 & 0.1137 & \multicolumn{1}{c}{---} & \multicolumn{1}{c}{---} \\
		~& c & 0.5259 & 0.5777 & 0.0904 & 0.0989 & 0.8778 & 0.1261 & \multicolumn{1}{c}{---} & \multicolumn{1}{c}{---} \\
		~& d & 0.5310 & 0.5665 & 0.0828 & 0.0774 & 0.8644 & 0.1130 & \multicolumn{1}{c}{---} & \multicolumn{1}{c}{---} \\
				
	    \multicolumn{1}{c}{\multirow{3}{*}{3200}} & a & 0.5165 & 0.5469 & 0.0859 & 0.1094 & 0.8125 & 0.0859 &
	         0.0547 & 0.8984 \\
	    ~& c & 0.5189 & 0.5441 & 0.0993 & 0.0963 & 0.8619 & 0.0971 & 0.0316 & 0.8689 \\
	    ~& d & 0.5186 & 0.5427 & 0.0827 & 0.1155 & 0.8272 & 0.0880 & 0.0532 & 0.9004 \\
	
	   \multicolumn{1}{c}{\multirow{3}{*}{5000}} & a & 0.5117 & 0.5352 & 0.0703 & 0.1367 & 0.8086 & 0.0742 &
	        0.0625 & 0.9102 \\
	   ~& b & 0.5176 & 0.5373 & 0.0784 & 0.1373 & 0.8078 & 0.0745 & 0.0667 & 0.9059 \\
	   ~& d & 0.5160 & 0.5375 & 0.0771 & 0.1303 & 0.8082 & 0.0760 & 0.0630 & 0.9146 \\
       \bottomrule
	\end{tabular}
	\begin{tablenotes}
	  \item{Note: a, Ghia et al.\cite{ghia1982high}; b, Hou et al.\cite{hou1994simulation}; c, Patil et al.\cite{patil2009finite}; d,    present work.}
	\end{tablenotes}
	\end{threeparttable}
\end{table}

\begin{table}
    \centering
	\caption{\label{tab:cavityeddyscales} The size of corner eddies at different Reynolds number.}
	\begin{tabular}{cccccccc}
	  \toprule
	    \multicolumn{1}{c}{\multirow{2}{*}{Re}} & \multicolumn{1}{c}{\multirow{2}{*}{Reference}} & \multicolumn{2}{c}{Lower left} &   \multicolumn{2}{c}{Lower right} &  \multicolumn{2}{c}{Upper left} \\
	    \cline{3-4}
	    \cline{5-6}
	    \cline{7-8}
	    ~&~& \multicolumn{1}{c}{$Width$} & \multicolumn{1}{c}{$Height$} & \multicolumn{1}{c}{$Width$} & \multicolumn{1}{c}{$Height$} &
	    \multicolumn{1}{c}{$Width$} & \multicolumn{1}{c}{$Height$} \\
	  \midrule
	    \multicolumn{1}{c}{\multirow{3}{*}{1000}} & Ghia et al.\cite{ghia1982high} & 0.2188 & 0.1680 & 0.3034 & 0.3536 &  \multicolumn{1}{c}{---} & \multicolumn{1}{c}{---} \\
	    ~& Patil et al.\cite{patil2009finite} & 0.2167 & 0.2005 & 0.2888 & 0.3431 & \multicolumn{1}{c}{---} & \multicolumn{1}{c}{---} \\
	    ~& Present & 0.2151 & 0.1641 & 0.2970 & 0.3530 & \multicolumn{1}{c}{---} & \multicolumn{1}{c}{---} \\
	
	    \multicolumn{1}{c}{\multirow{3}{*}{3200}} & Ghia et al.\cite{ghia1982high} & 0.2844 & 0.2305 & 0.3406 & 0.4102 & 0.0859 & 0.2057 \\
	    ~& Patil et al.\cite{patil2009finite} & 0.3146 & 0.2443 & 0.3018 & 0.3925 & 0.0494 & 0.1819 \\
	    ~& Present & 0.2816 & 0.2356 & 0.3358 & 0.3839 & 0.0877 & 0.1938 \\
	
	    \multicolumn{1}{c}{\multirow{2}{*}{5000}} & Ghia et al.\cite{ghia1982high} & 0.3184 & 0.2643 & 0.3565 & 0.4180 & 0.1211 & 0.2693 \\
	     ~& Present & 0.3013 & 0.2531 & 0.3432 & 0.3725 & 0.1178 & 0.2367 \\
	  \bottomrule
	\end{tabular}
\end{table}

\subsection{Laminar flows around a cylinder}
\begin{figure}
  	\centering
  	\subfigure{
  			\includegraphics[width=0.5 \textwidth]{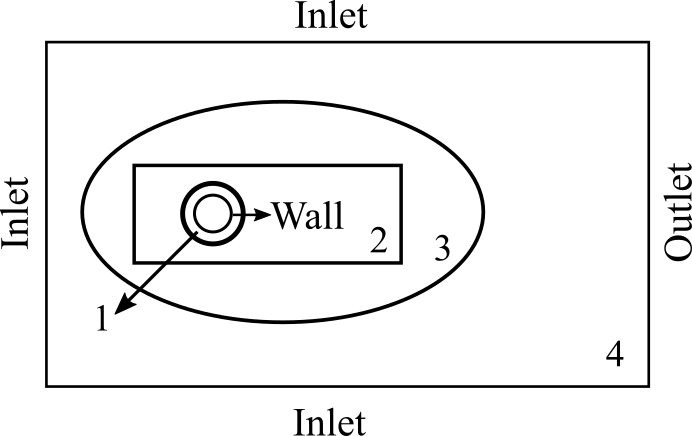}
  		}
  	\caption{\label{cylinderconfig} Configuration for flow around single circular cylinder, boundary conditions and regions for grid refinement ($1,2,3,4$ represent different grid refined regions, respectively).}
  \end{figure}

 For the flow around the single circular cylinder, depending on the Reynolds number, the flow will present steady flow or unsteady flow. Two counter-rotating vortexes are presented if flow is steady and $Re>1$ and periodic vortex shed, named Karman Vortex Streets, are present if flow is unsteady. Fig.~\ref{cylinderconfig} shows the configure of computational domain, boundary conditions and different regions for grid refinement. The strategy to generate this grid is refined successive from far-field to near wall region. The near wall domain (region $1$ in Fig.~\ref{cylinderconfig}) can be refined use triangles or quadrangles to enhance the computational accuracy, and the refined range for wake after cylinder (region $2$ in Fig.~\ref{cylinderconfig}) can adjustable depending on the Reynolds number. First, the same test in Ref.~\cite{patil2012two} with $Re = 100$ is adopted to validate the present method can apply to unsteady flow. The computational domain is $70d \times 50d$ and the cylinder is set into the center location of domain, where the $d$ is the cylinder diameter. Two grids are used in this test, which the region $1$ are refined by triangles and quadrangles, respectively, and other regions use the same grid (all discretized into triangle). The width of region $1$ is $0.2d$, the the circumference of the cylinder is discrete into $360$ points, the amount of grid cells in this domain are $8640$ and $6992$, respectively, and the size of first cell near the wall are about $0.0015d$ and $0.008d$, respectively. About $10d$ length is refined (region $2$) to capture the shed vortexes. Finally, the total amount cells of two grids are $106157$ (full unstructured grid) and $108165$ (hybrid grid), respectively. For the boundary conditions, the left, top and bottom border are set into the inlet boundary condition, the right border is set into outlet boundary condition, and the cylinder is set into wall boundary condition. The density $\rho_{\infty}=1.0$ and velocity $u_{\infty}=0.1, v_{\infty}=0.0$ are used for the initial condition. The Reynolds number for this flow is defined as $Re = u_{\infty}d/{\nu}$.

\begin{figure}
 	\centering
 	\subfigure[]{
 			\includegraphics[width=0.45 \textwidth]{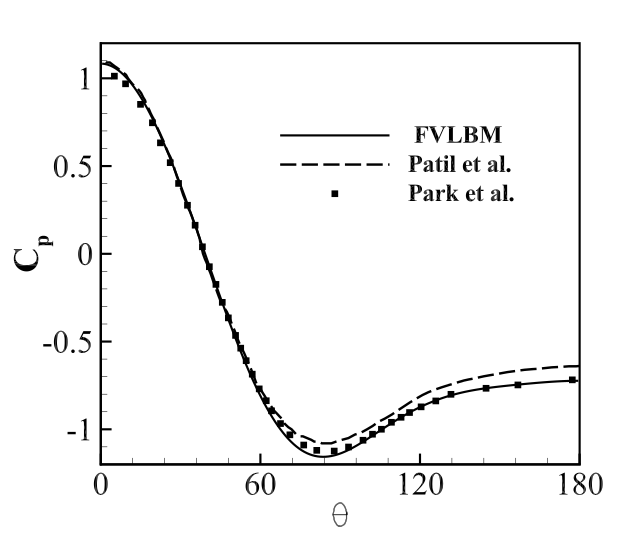}
 		}
    \subfigure[]{
     		\includegraphics[width=0.45 \textwidth]{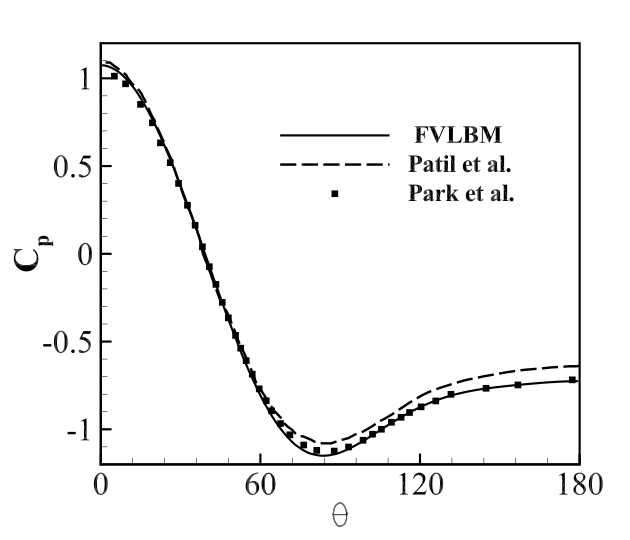}
     		}
 	\caption{\label{CpRe100compare} Pressure coefficient distributions for flow around circular cylinder at $Re = 100$ for (a) hybrid grids and (b) full triangle unstructured grid.}
 \end{figure}

 The pressure coefficient is defined as

 \begin{equation}
           C_p = \frac{p-p_{\infty}}{0.5\rho_{\infty}{u^2_{\infty}}},
 \end{equation}
 where $p$ is the pressure at the wall, and is approximated with the value at the center of the first cell near the wall. Fig.~\ref{CpRe100compare} shows the comparisons of pressure coefficient, it is clear that the difference between two grids is little and we have good results compared with Park et al.'s. In additional, the good result obtained on hybrid grid illustrate the advantage of present method, which can decline the amount of grid cells to simulate the hight Reynolds number flows in some conditions.

\begin{figure}
	\centering
	\subfigure[]{
			\label{cylindermesh:a}
			\includegraphics[width=0.6 \textwidth]{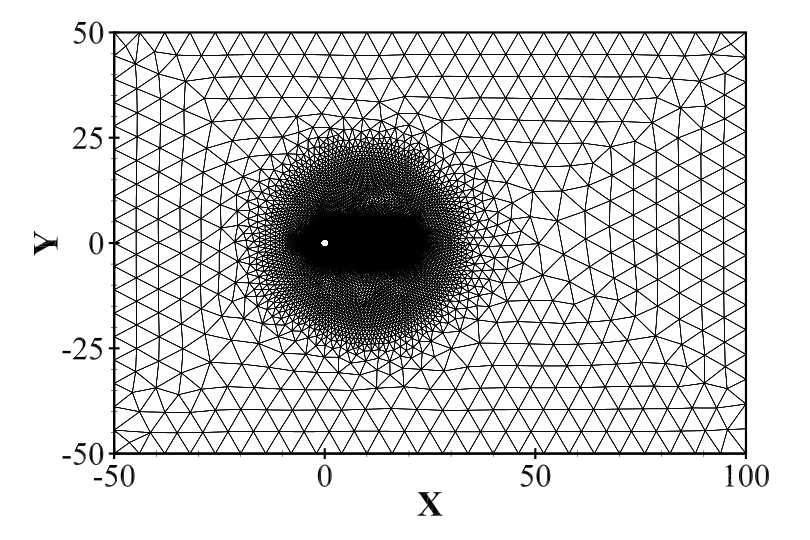}
		}
    \subfigure[]{
    		\label{cylindermesh:b}
    		\includegraphics[width=0.6 \textwidth]{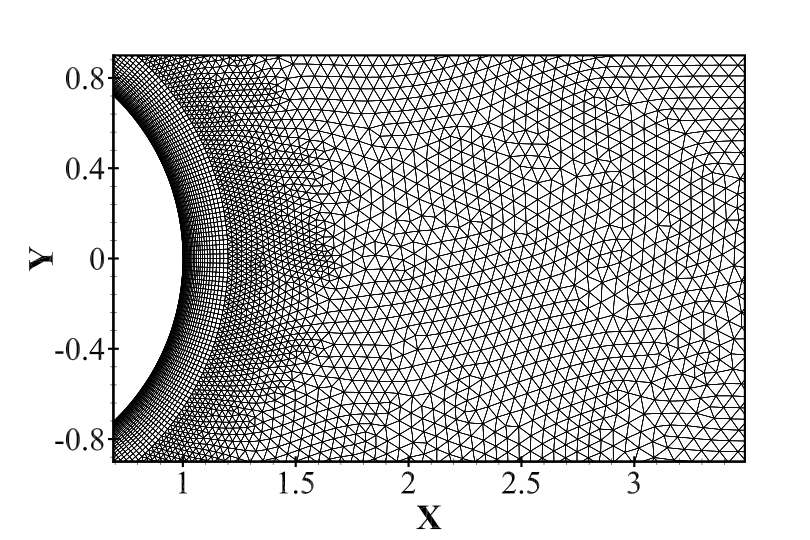}
    	}
	\caption{\label{cylindermesh} Grid for flow around circular cylinder: (a) Full domain and (b) near the cylinder surface.}
\end{figure}

 Next, we simulate the flows at different Reynolds numbers, from steady flow to unsteady flows. In this study, the initial conditions and boundary conditions are same as to above. The grid also hybrid and the computational domain is reset into a $75d\times50d$ rectangle, the location of cylinder is $(25d,25d)$. The computational domain is very large so as to eliminate the influence of the far-field boundary condition.  Fig.~\ref{cylindermesh} shows the hybrid unstructured grid used in this study. The total number of cells is 108179, where the quadrilateral cell is 8640, and others is the triangular cell. $Re = 10$, $20$, $40$, $45$, $46$, $47$, $60$, $80$ and $100$ is considered in the simulation. For the temporal discretized, the Euler scheme is used at $Re = 10,20,40$, and AB2 scheme is used at other Reynolds number. The time step is set to $0.001$ for all simulations.

\begin{figure}
	\centering
	\subfigure[]{
			\label{cylinderRe10}
			\includegraphics[width=0.45 \textwidth]{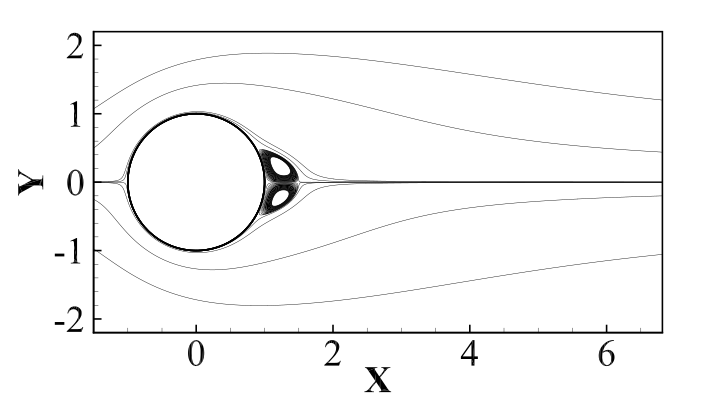}
		}
    \subfigure[]{
			\label{cylinderRe20}
			\includegraphics[width=0.45 \textwidth]{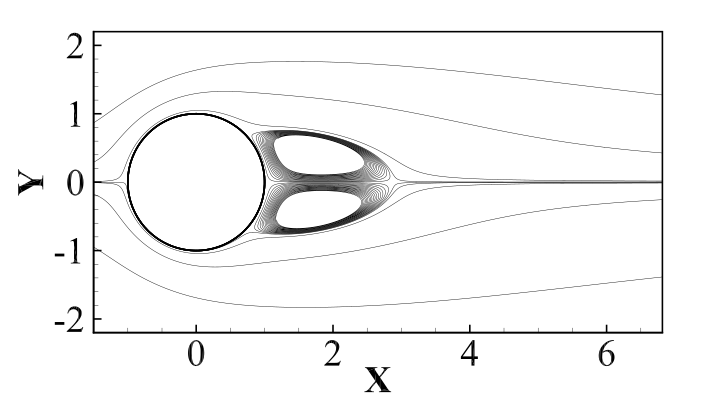}
		}
    \subfigure[]{
			\label{cylinderRe40}
			\includegraphics[width=0.45 \textwidth]{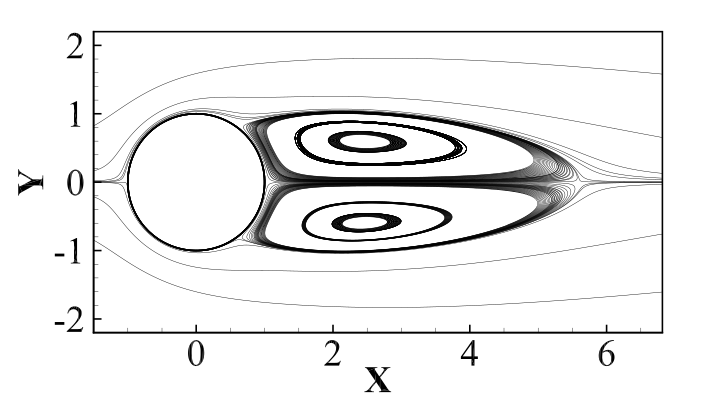}
		}
    \subfigure[]{
			\label{cylinderRe60}
			\includegraphics[width=0.45 \textwidth]{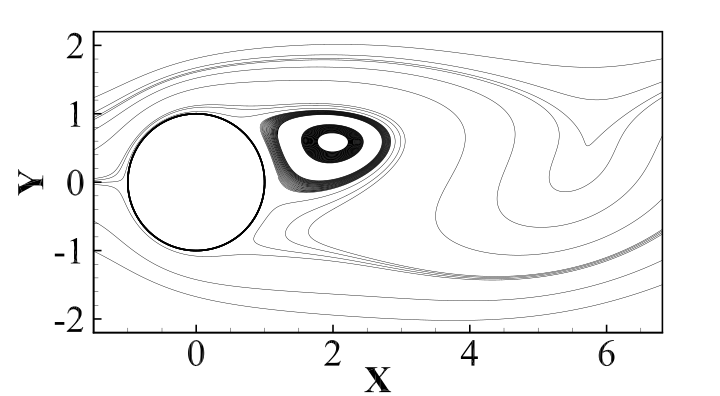}
		}		
	\caption{\label{cylinderstreamlines} Streamlines around circular cylinder at (a) $Re = 10$, (b) $Re = 20$, (c) $Re = 40$ and (d) $Re = 60$.}
\end{figure}

Fig.~\ref{cylinderstreamlines} shows the streamlines around the cylinder at different Reynolds number. As can be observed in these figures, for the case $Re \leq 40$, the flow is steady and the length of separate vortex is increase with the enhancement of Reynolds number, and for $Re = 60$, the flow is unsteady.

\begin{figure}
	\centering
	\subfigure[]{
			\label{cdsteady}
			\includegraphics[width=0.45 \textwidth]{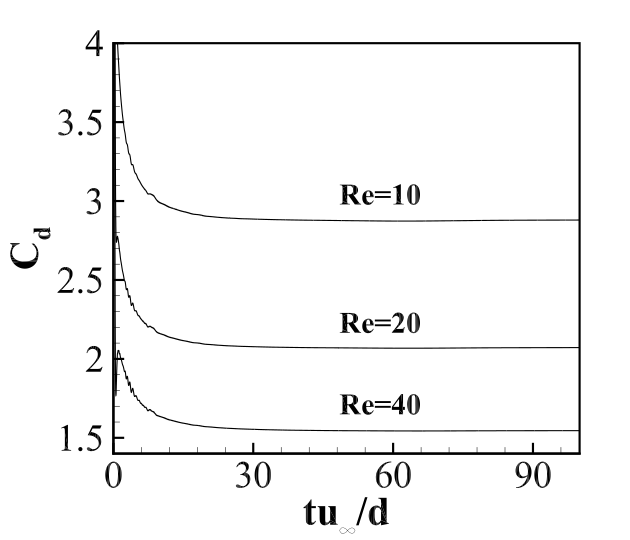}
		}
    \subfigure[]{
    		\label{cdunsteady}
    		\includegraphics[width=0.45 \textwidth]{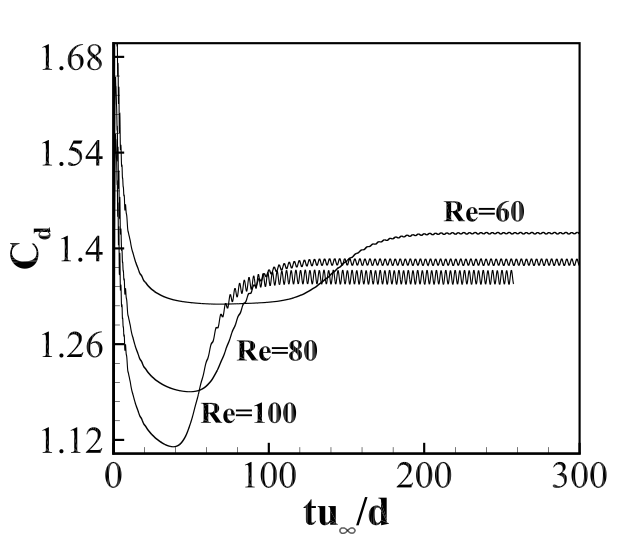}
    	}
    \subfigure[]{
            \label{clRe40}
            \includegraphics[width=0.45 \textwidth]{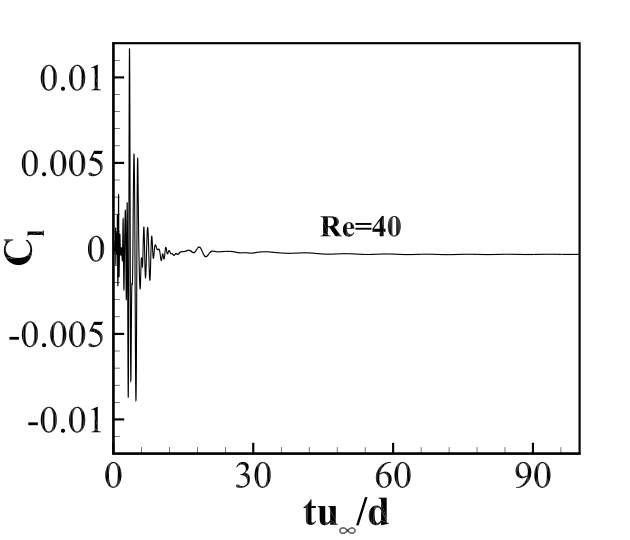}
        }
    \subfigure[]{
        	\label{cl}
       		\includegraphics[width=0.45 \textwidth]{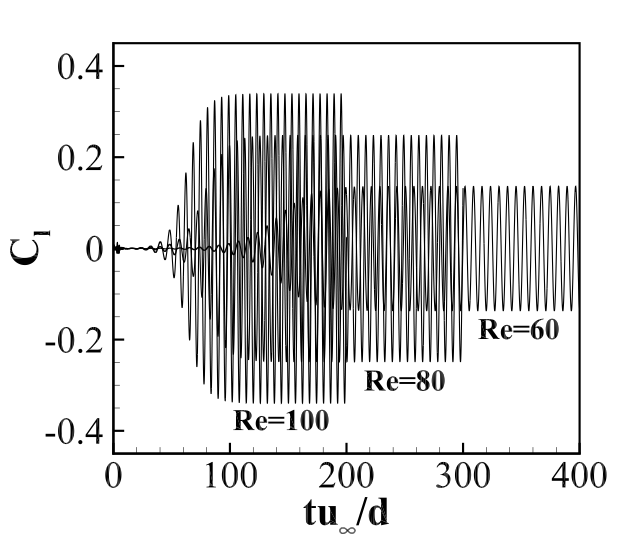}
        }
	\caption{\label{cylinderCdCl} Time evolutions of (a)(b) the drag coefficient and (c)(d) the lift coefficient for flow around circular cylinder at different Reynolds number.}
\end{figure}

\begin{figure}
	\centering
	\subfigure[]{
			\label{cylinderRe45}
			\includegraphics[width=0.45 \textwidth]{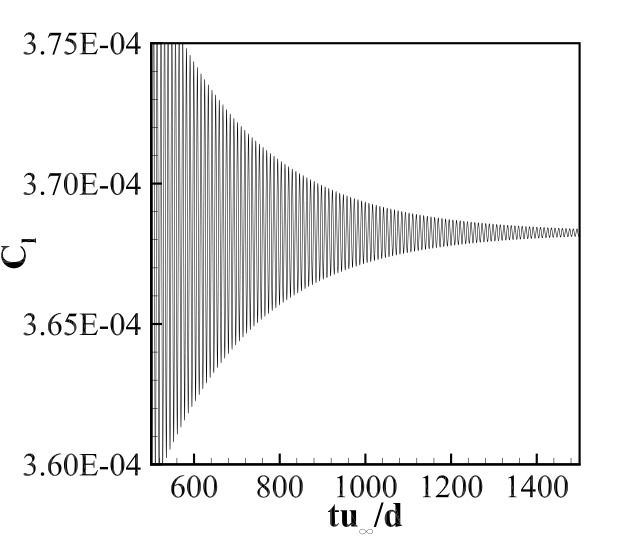}
		}
    \subfigure[]{
    		\label{cylinderRe47}
    		\includegraphics[width=0.45 \textwidth]{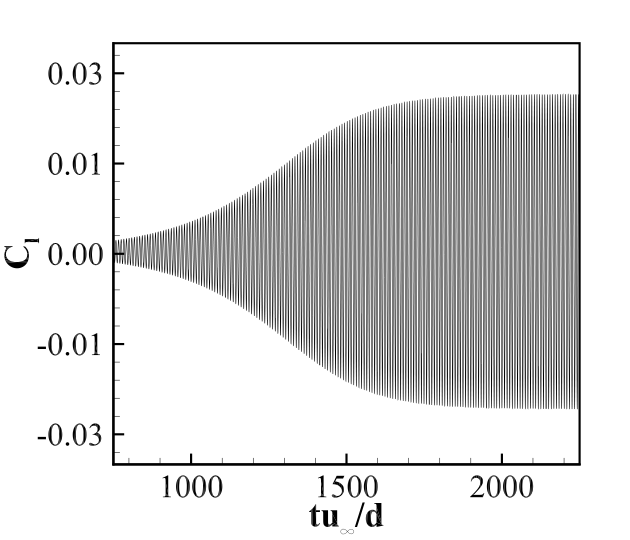}
    	}
	\caption{\label{ClRe45-47} Time evolutions of the lift coefficient for flow around circular cylinder at (a) $Re = 45$ and (b) $Re = 47$.}
\end{figure}

\begin{figure}
	\centering
	\subfigure[]{
			\label{cylinderstreamlinesRe45}
			\includegraphics[width=0.4 \textwidth]{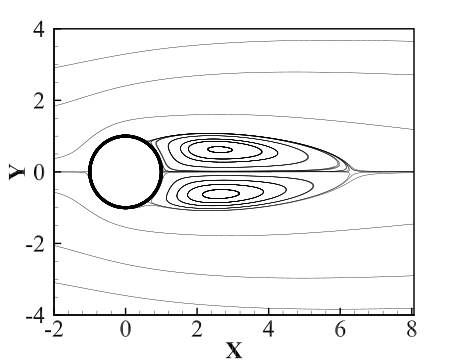}
			}
	\subfigure[]{
			\label{cylinderstreamlinesRe47}
			\includegraphics[width=0.478 \textwidth]{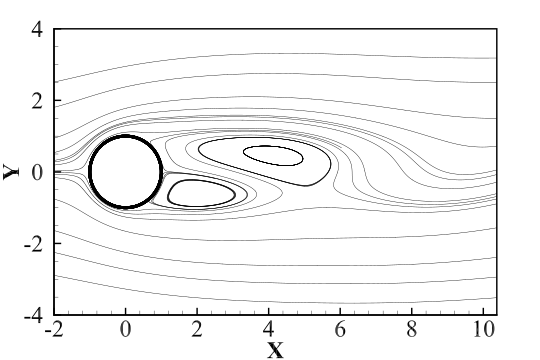}
		}
	\caption{\label{cylinderstreamlinesRe45-47} The streamlines of flow around circular cylinder at (a) $Re = 45$ and (b) $Re = 47$.}
\end{figure}

\begin{figure}
	\centering
	\subfigure[]{
	        \label{fullcurveRe46}
			\includegraphics[width=0.45 \textwidth]{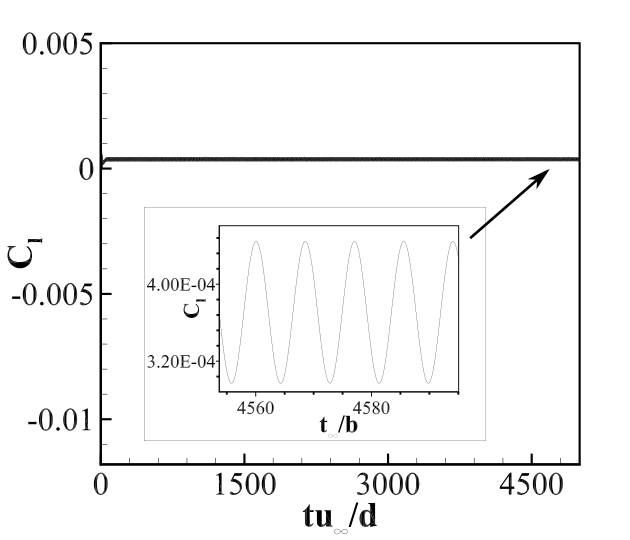}
		}
    \subfigure[]{
            \label{amplitudeRe46}
    		\includegraphics[width=0.45 \textwidth]{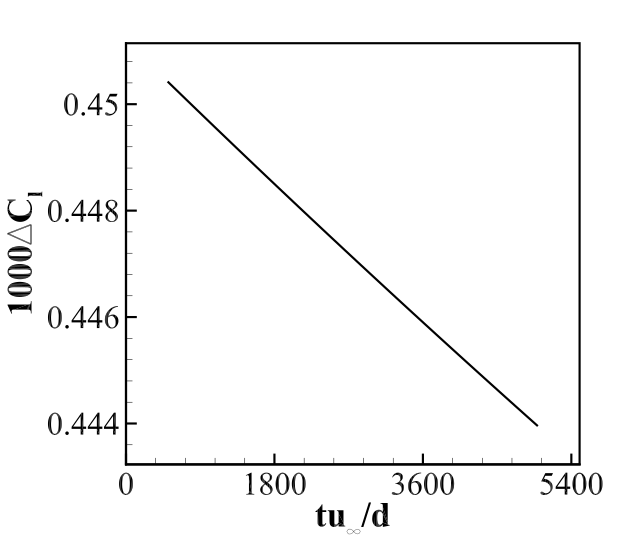}
    	}
	\caption{\label{ClRe46} Time evolution of the lift coefficient for flow around circular cylinder at $Re = 46$ with (a) full and partial of curve and (b) the change of the amplitude of curve.}
\end{figure}

The lift and drag coefficient are defined as

\begin{equation}
      \begin{aligned}
         &  C_l = \frac{F_l}{0.5\rho{u_{\infty}^2}d}, \\
         &  C_d = \frac{F_d}{0.5\rho{u_{\infty}^2}d},
      \end{aligned}
\end{equation}
where $F_l$ and $F_d$ are the lift and drag force, respectively, and can be calculated use the equation $(3.10)$ in Ref.~\cite{lifinite}. Fig.~\ref{cylinderCdCl} shows the time evolutions of $C_l$ and $C_d$ at different Reynolds number, where $tu_{\infty}/d$ is the dimensionless time. It's clear that for $Re\leq40$ the flow is steady, as the $C_d$ can converges to a constant (see Fig.~\ref{cdsteady}) and $C_l$ also can converges to a constant that close to 0 (see Fig.~\ref{clRe40}). For $Re\geq60$, the flow is unsteady, because both $C_l$ and $C_d$ will present steady sinusoidal fluctuation with advance of the simulation time to some critical value (see Fig.~\ref{cdunsteady}-\ref{cl}). To estimate the performance  of the FVLBM present in this paper at critical Reynolds number $Re_{cr}$ for the vortex shedding, we first simulate the flow at $Re = 45, 47$. Fig.~\ref{ClRe45-47} shows the time evolutions of $C_l$ at $Re = 45,47$. It can be observed that the curve of lift coefficient for $Re = 45$ is convergence obviously and for $Re = 47$ is develop into sinusoidal fluctuation. It means that the flow is steady at $Re = 45$ and is unsteady at $Re = 47$. This conclusion also can be confirmed from the streamlines around the cylinder at two Reynolds numbers in Fig.~\ref{cylinderstreamlinesRe45-47}. So the $Re_{cr}$ is $45<{Re_{cr}}<47$. To decrease the range of $Re_{cr}$, we next simulate the flow at $Re = 46$. For $Re = 46$, it can be found that the curve of lift coefficient present periodic fluctuation from Fig.~\ref{fullcurveRe46}. To estimate if the curve is convergence, the change of the amplitude of curve is calculated (see Fig.~\ref{amplitudeRe46}). We can find the flow is also steady as the amplitude of curve of $C_l$ is convergence, although the convergence is very slow and will take a much long time. So, the $Re_{cr}$ is between $46$ and $47$. This value is same as to linear stability theory and numerous numerical and experiment values\cite{Yuan2015An}.

The length of separate vortex is another important parameter to validate the numerical method. For the unsteady flow, as the flow is change periodic, the $L$ is obtained from the time-averaged variables of the flow field. The dimensionless vortex shedding frequency can be defined as

\begin{equation}
        St = \frac{fd}{u_\infty},
\end{equation}
where $St$ is Strouhal number, and $f$ is the vortex shedding frequency. The values of $C_d$, $St$, and $L$ are present in Tables \ref{tab:cylinderRe10-20-40} and\ref{tab:cylinderRe60-80-100}. For $Re\geq47$, the $C_d$ is also calculated from time-averaged variables. The experiment values presented by Tritton\cite{tritton1959experiments} are obtained from the fitting curve based on the experiment values. It can be found that the mean drag coefficient obtained from FVLBM is little higher than experiment values and other numerical methods at $Re\geq47$. In additional, compared with parallel vortex shedding mode\cite{williamson1989oblique}, which the experimental fitting curve  of $St$ can be expressed as $St=-3.3265/Re+0.1816+0.00016Re$, the max error for these three Reynolds numbers is about $1.62\%$.

\begin{table}\tiny
    \centering
	\caption{\label{tab:cylinderRe10-20-40} Comparison of drag coefficient $C_d$ and length of separate vortex $L$ for flow around circular cylinder at $Re = 10,20,40$.}
	\begin{tabular}{cccccccccccc}
	  \toprule
	    \multicolumn{1}{c}{\multirow{2}{*}{Re}} &  \multicolumn{1}{c}{Tritton\cite{tritton1959experiments}} &
	    \multicolumn{2}{c}{Fornberg\cite{fornberg1980numerical}}  & \multicolumn{2}{c}{Park et al.\cite{park1998numerical}} & \multicolumn{2}{c}{He et al.\cite{he1997lattice}} &  \multicolumn{2}{c}{Li et al.\cite{lifinite}}  &
	    \multicolumn{2}{c}{Present} \\
	    \cline{2-2}
	    \cline{3-4}
	    \cline{5-6}
	    \cline{7-8}
	    \cline{9-10}
	    \cline{11-12}
	    ~& \multicolumn{1}{c}{$C_d$} & \multicolumn{1}{c}{$C_d$} & \multicolumn{1}{c}{$L/R$} &
	       \multicolumn{1}{c}{$C_d$} & \multicolumn{1}{c}{$L/R$} & \multicolumn{1}{c}{$C_d$} & \multicolumn{1}{c}{$L/R$} &
	       \multicolumn{1}{c}{$C_d$} & \multicolumn{1}{c}{$L/R$} & \multicolumn{1}{c}{$C_d$} & \multicolumn{1}{c}{$L/R$} \\
	  \midrule
	    \multicolumn{1}{c}{10} & 2.926 & \multicolumn{1}{c}{---} & \multicolumn{1}{c}{---} & 2.78 &  0.476   &
	    3.170 & 0.474 &	3.003 & 0.6649 & 2.88 & 0.502  \\
        \multicolumn{1}{c}{20} & 2.103 & 2.0001 & 1.82 & 2.01 &  1.814   &
	    2.152 & 1.842 &	2.118 & 2.0376 & 2.072 & 1.866 \\
	    \multicolumn{1}{c}{40} & 1.605 & 1.4980 & 4.48 & 1.51 &  4.502   &
	    1.499 & 4.490 &	1.568 & 4.7027 & 1.545 & 4.609 \\	
	  \bottomrule
	\end{tabular}
\end{table}

\begin{table}\tiny
    \centering
	\caption{\label{tab:cylinderRe60-80-100} Comparison of mean drag coefficient $C_d$, length of separate vortex $L$ and Strouhal number $St$ for flow around circular cylinder at $Re = 60,80,100$.}
	\begin{tabular}{ccccccccccc}
	  \toprule
	    \multicolumn{1}{c}{\multirow{2}{*}{Re}} &  \multicolumn{1}{c}{Tritton\cite{tritton1959experiments}} &
	    \multicolumn{1}{c}{Fornberg\cite{fornberg1980numerical}}  & \multicolumn{3}{c}{Park et al.\cite{park1998numerical}} &
	    \multicolumn{2}{c}{Zarghami et al.\cite{zarghami2012lattice}} &
	    \multicolumn{3}{c}{Present} \\
	    \cline{2-2}
	    \cline{3-3}
	    \cline{4-6}
	    \cline{7-8}
	    \cline{9-11}
	    ~& \multicolumn{1}{c}{$C_d$} & \multicolumn{1}{c}{$C_d$} &
	       \multicolumn{1}{c}{$C_d$} & \multicolumn{1}{c}{$L/R$} & \multicolumn{1}{c}{$St$} & \multicolumn{1}{c}{$C_d$} &
	       \multicolumn{1}{c}{$St$}  & \multicolumn{1}{c}{$C_d$} &
	       \multicolumn{1}{c}{$L/R$} & \multicolumn{1}{c}{$St$}  \\
	  \midrule
	    \multicolumn{1}{c}{60} & 1.398 & \multicolumn{1}{c}{---} & 1.39 & 4.132 &  0.1353   &
	    \multicolumn{1}{c}{---} & \multicolumn{1}{c}{---} &
	     1.422 & 4.155 & 0.1375 \\
	    \multicolumn{1}{c}{80} & 1.316 & \multicolumn{1}{c}{---} & 1.35 & 3.312 &  0.1528   &
	    \multicolumn{1}{c}{---} & \multicolumn{1}{c}{---}  &
	    1.379 & 3.306 & 0.1550 \\
	    \multicolumn{1}{c}{100} & 1.271 & 1.058 & 1.33 & 2.782 &  0.1646   &
	    1.310 & 0.161 &
	    1.358 & 2.796 & 0.1670 \\
	  \bottomrule
	\end{tabular}
\end{table}

\begin{figure}
	\centering
	\subfigure[]{
			\label{cylinderCp10-20-40}
			\includegraphics[width=0.45 \textwidth]{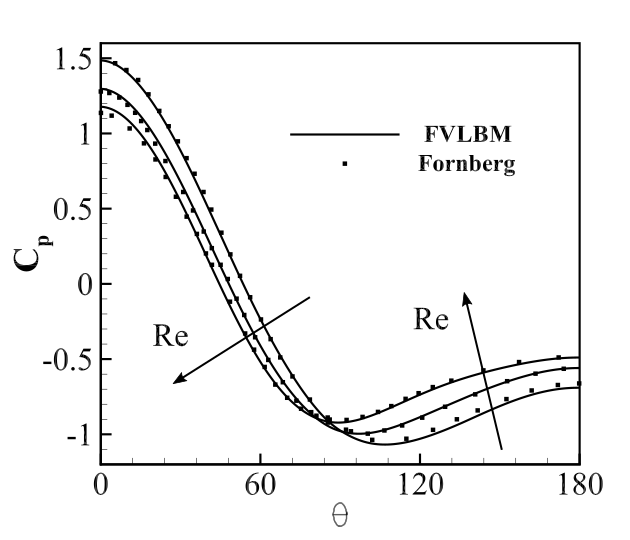}
			}
    \subfigure[]{
			\label{cylinderCp60-80-100}
			\includegraphics[width=0.45 \textwidth]{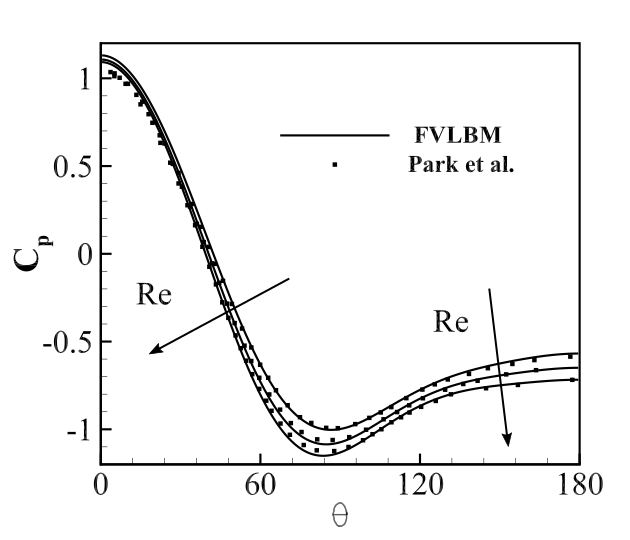}
			}
	\caption{\label{cylinderCp} Pressure coefficient distributions for flow around circular cylinder at (a) $Re = 10,20,40$ and (b) $Re = 60,80,100$ compared with Fornberg\cite{park1998numerical} and Park et al.\cite{fornberg1980numerical}.}
\end{figure}

Fig.~\ref{cylinderCp} shows the $C_p$ around the wall of cylinder and compared with Ref.~\cite{fornberg1980numerical,park1998numerical}, it can be found though the $C_p$ at leading stagnation point is little higher, the FVLBM can obtain good results compared with other numerical methods.

\subsection{Laminar flows around double circular cylinders}
 The last test case in this paper is the flow around double circular cylinders. Cylinder-like structures have many applications such as in the designs for buildings, chimneys, heat exchangers and so on. Compared to the flow around single circular cylinder, the studies of flow around bi-cylinder are fewer. A review for recent studied can be found in Ref.~\cite{sumner2010two}. For the flow around multiple-cylinder, the most simple examples are the flow around two cylinders with equal diameter arranged in tandem and side-by-side. The flow patterns for these flows depended on the Reynolds number $Re$ and the spacing $s$ between two cylinders, where the dimensionless cylinder spacing $s$ is defined as $s = l/d$, $l$ is the dimension distance between the center of two cylinders, and $d$ is the dimension diameter of cylinder. Here, the $Re = 100$ is chose for two arrangements of cylinders as the flow at this Reynolds number is laminar flow.

\subsubsection{Arranged in tandem}
\begin{figure}
 	\centering
 	  \subfigure[]{
 			\includegraphics[width=0.51 \textwidth]{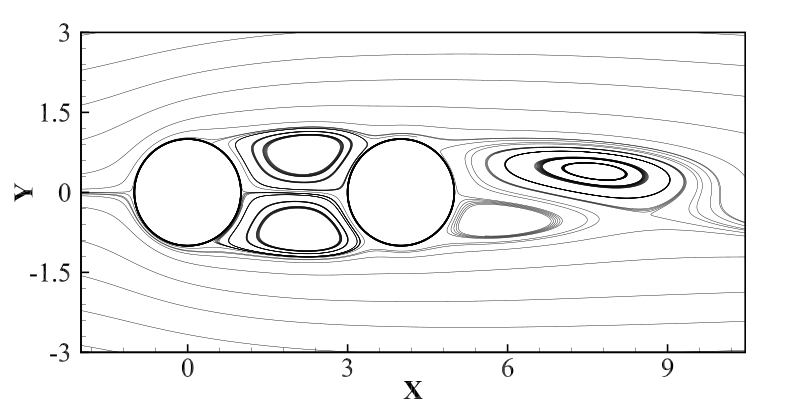}
 			\includegraphics[width=0.45 \textwidth]{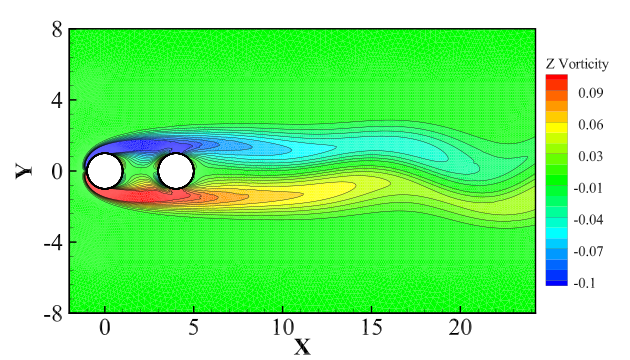}
 			}
 	  \subfigure[]{
 	   		\includegraphics[width=0.51 \textwidth]{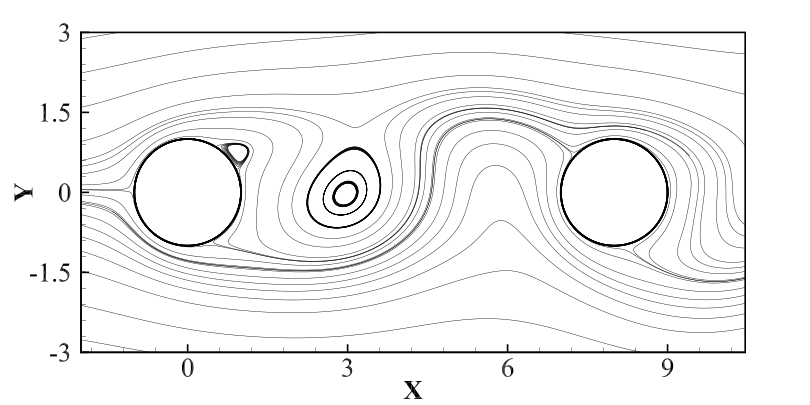}
 	   		\includegraphics[width=0.45 \textwidth]{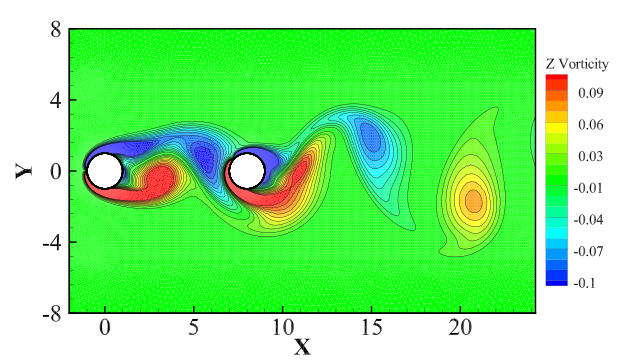}
 	   		}
 	\caption{\label{cylindertandemstream} Instantaneous streamlines (left) and vorticity contours (right) for flow over two cylinders arranged in tandem at $Re = 100$ with spacings (a) $s = 2$ and (b) $s = 4$.}
\end{figure}

For the flow around two cylinders arranged in tandem, these exist three flow patterns as the changing of spacing $s$: (a) extended-body regime, (b) reattachment regime and (c) co-shedding regime\cite{xu2004strouhal}. According to Sharman et al.\cite{sharman2005numerical}, the critical spacing $s_c$ that flow pattern transit from (b) to (c) is about 3.75-4. Thus, in this section, two spacings, $s = 2$ and $s = 4$ are involved in our simulation to obtain two different flow patterns. The size of computational domain, initial and boundary conditions are same as to the test case of flow around single circular cylinder. The upstream cylinder is located at $(25d,25d)$, and the total number of cells of the hybrid grids at two spacings are 118062 and 119102 (less than the number of grid cells used in Ref.~\cite{patil2012two}), respectively. Fig.~\ref{cylindertandemstream} shows the instantaneous streamlines and vorticity contours at two spacings. It can be found that at spacing $s = 2$, two counter-rotating vortexes appeared in the gap and can not shed from the upstream cylinder as the suppresses of downward cylinder, and at spacing $s = 4$, the vortexes shed from each cylinder.

\begin{figure}
 	\centering
 	  \subfigure[]{
 			\includegraphics[width=0.45 \textwidth]{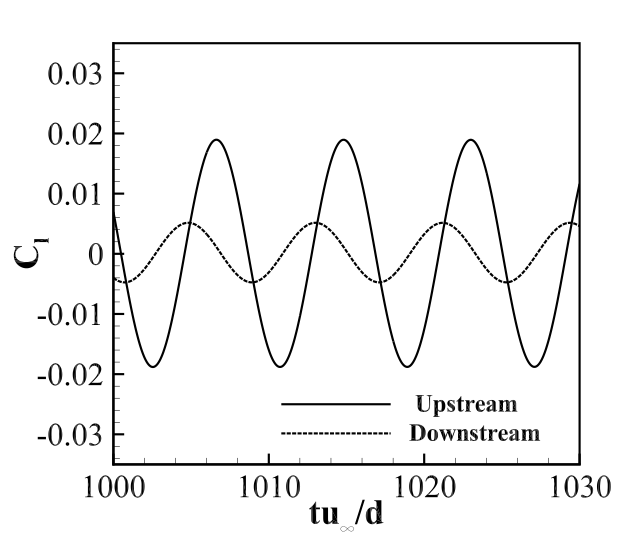}
 			}
 	  \subfigure[]{
 	   		\includegraphics[width=0.45 \textwidth]{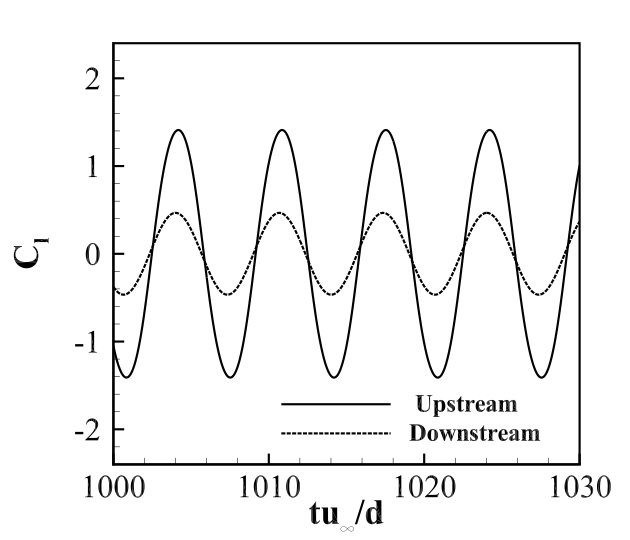}
 	   		}
 	\caption{\label{cylindertandemCl} Time evolutions of the lift coefficient for flow around two cylinders arranged in tandem at $Re = 100$ with spacings (a) $s = 2$ and (b) $s = 4$.}
\end{figure}

The definitions of the lift coefficient $C_l$, the drag coefficient $C_d$, the pressure coefficient $C_p$ and Strouhal number $St$ are same as to the above. Both $C_l$ and $C_d$ of two cylinders will develop into steady sinusoidal fluctuation at two different spacing and share a same frequency (see Fig.~\ref{cylindertandemCl}), these results are consist with previous works\cite{sumner2010two,sharman2005numerical}. For the curve of the $C_l$, the phases of two cylinders at spacing $s = 2$ are almost in-phase but have phase difference at spacing $s = 4$. For the distribution of $C_p$ at upstream cylinder, the difference between ours and Sharman et al.'s is little at leading stagnation but lower than latter at trailing stagnation for both spacings. For the downstream cylinder, the shape of distribution is consist with Sharman et al.'s but translate downward some values which same as to the difference at trailing stagnation of upward cylinder. It means that the fluctuation of pressure near the upward cylinder will propagate to downward cylinder. Patil et al. obtain the same results at spacing $s = 2$ but higher than Sharman et al.'s. As without more reference, we can not identify which result is better. Table \ref{tab:cylindertandem} is some compared results with other studies, where $\tilde{C}_l$ and $\tilde{C}_d$ are the root-mean-square drag and lift coefficients, respectively. The root-mean-square coefficients of Mizushima et al.\cite{mizushima2005instability} in Table \ref{tab:cylindertandem} are calculated through dividing the amplitudes by $\sqrt{2}$ according to the sinusoidal-fluctuation approximation. The $C_d$ of upward cylinder all have positive values at two spacings, the $C_d$ of downward cylinder has negative value at spacing $s = 2$ and has positive value at spacing $s = 4$. Compared with other works, our results are in reasonable range. Besides, for $St$ we have good results compared with Sharman et al.\cite{sharman2005numerical}.

\begin{figure}
 	\centering
 	   \subfigure[]{
 			\includegraphics[width=0.45 \textwidth]{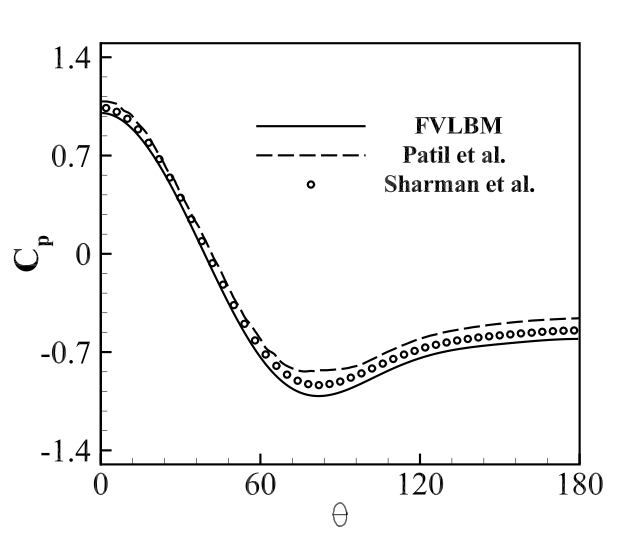}
 			\includegraphics[width=0.45 \textwidth]{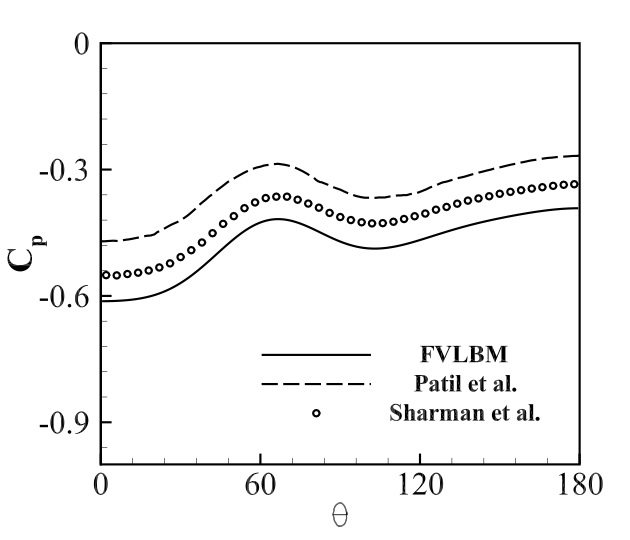}
 			}
 	   \subfigure[]{
 	    	\includegraphics[width=0.45 \textwidth]{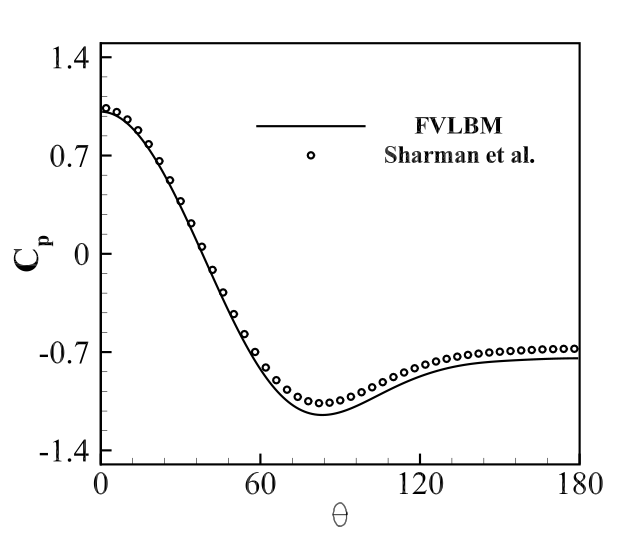}
 			\includegraphics[width=0.45 \textwidth]{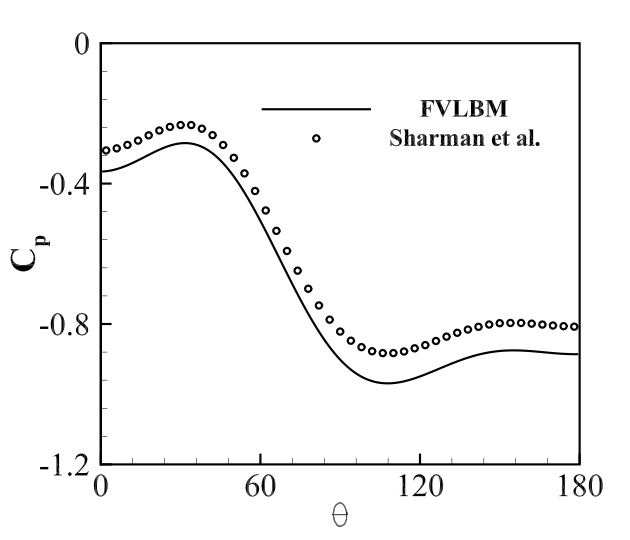}
 	    	}
 	\caption{\label{cylindertandemCp} Mean pressure coefficient distributions for flow around two cylinders arranged in tandem with spacings (a) $s = 2$ and (b) $s = 4$ on the upstream cylinder (left) and the downstream cylinder (right) at $Re = 100$.}
\end{figure}

\begin{table}\tiny
    \centering
	\caption{\label{tab:cylindertandem} The drag coefficient, lift coefficient and Strouhal number for flow around two cylinders arranged in tandem.}
	\begin{tabular}{ccccccccc}
	  \toprule
	    \multicolumn{1}{c}{\multirow{2}{*}{Reference}} &  \multicolumn{1}{c}{\multirow{2}{*}{$s$}} &
	    \multicolumn{3}{c}{Upstream cylinder}  & \multicolumn{3}{c}{Downstream cylinder} &
	    \multicolumn{1}{c}{\multirow{2}{*}{$St$}} \\
	    \cline{3-5}
	    \cline{6-8}
	    ~&~& \multicolumn{1}{c}{$\overline{C}_d$} & \multicolumn{1}{c}{$\tilde{C}_d$} &
	       \multicolumn{1}{c}{$\tilde{C}_d$} & \multicolumn{1}{c}{$\overline{C}_d$} & \multicolumn{1}{c}{$\tilde{C}_d$} & \multicolumn{1}{c}{$\tilde{C}_d$} &~ \\
	  \midrule
	    \multicolumn{1}{c}{Sharman et al.\cite{sharman2005numerical}} & 2 & 1.168 & 0.000 & 0.007 & -0.088 &  0.000   &
	    0.027 & 0.123 \\
	    \multicolumn{1}{c}{Mizushima et al.\cite{mizushima2005instability}} & 2 & 1.196 & \multicolumn{1}{c}{---} & 0.024 &
	    -0.043 &  \multicolumn{1}{c}{---} & 0.006 & 0.125 \\
	    \multicolumn{1}{c}{Patil et al.\cite{patil2012two}} & 2 & 1.1369 & \multicolumn{1}{c}{---} & 0.0013 & -0.0669 &
	          \multicolumn{1}{c}{---} & 0.0039 & 0.1157 \\
	    Present & 2 & 1.187 & 0.000 & 0.003 & -0.091 & 0.000 & 0.013 & 0.122 \\
	    \multicolumn{1}{c}{Sharman et al.\cite{sharman2005numerical}} & 4 & 1.277 & 0.016 & 0.303 & 0.707 & 0.141 & 0.987 & 0.148 \\
	    \multicolumn{1}{c}{Mizushima et al.\cite{mizushima2005instability}} & 4 & 1.327 & \multicolumn{1}{c}{---} & 0.271 & 1.016 &
           \multicolumn{1}{c}{---} & 1.064 & 0.160 \\
        Papaioannou et al.\cite{papaioannou2006three} & 4 & 1.31 & \multicolumn{1}{c}{---} & \multicolumn{1}{c}{---} & 0.75 &
           \multicolumn{1}{c}{---} & \multicolumn{1}{c}{---} & 0.152 \\
        Present & 4 & 1.304 & 0.023 & 0.329 & 0.728 & 0.149 & 1.015 & 0.149 \\
	  \bottomrule
	\end{tabular}
\end{table}

\subsubsection{Arranged side-by-side}
For the flow around two cylinders arranged side-by-side, it can be divided into three main flow patterns: (a) single-bluff-body pattern, (b) a biased flow pattern and (c) parallel vortex streets shedding pattern\cite{sumner2010two}. The value of spacing $s$ that transit from pattern (b) to (c) is about 2-2.2. In this section, $s =2$ and $s =2.5$ are chose to simulate two different flow patterns. The size of computational domain, initial and boundary conditions are same as to the test case of flow around single circular cylinder. The centers of two cylinders are placed symmetric in $y$-direction, where the symmetrical point is set at $(25d,25d)$. The total number of cells of hybrid grids for two spacings are 120850 and 136134 (less than the number of grids used in Ref.~\cite{patil2012two}), respectively.

\begin{figure}
 	\centering
 	\subfigure[]{
 			\label{cylinderClsidesides2}
 			\includegraphics[width=0.45 \textwidth]{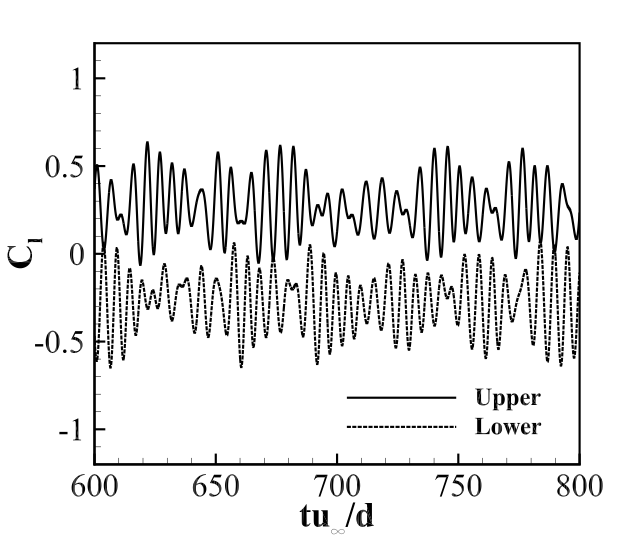}
 			}
     \subfigure[]{
 			\label{cylinderCdsidesides2}
 			\includegraphics[width=0.45 \textwidth]{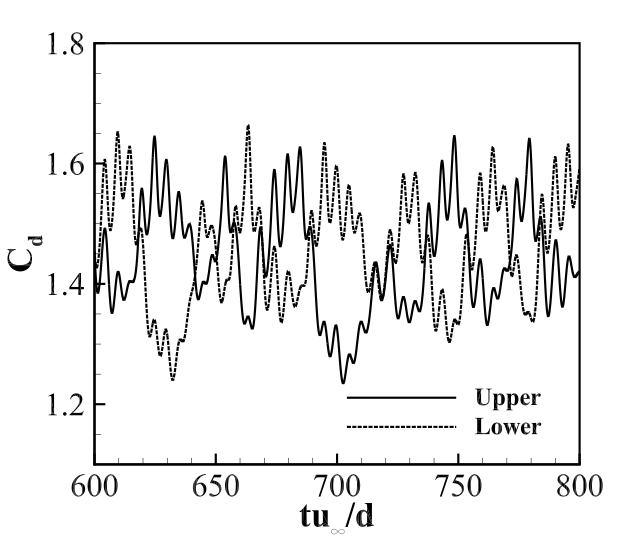}
 			}
 	\caption{\label{cylinderClCdsidesides2} Time evolutions of (a) the lift coefficient and (b) the drag coefficient for flow around two cylinders arranged side-by-side with spacing $s = 2$ at $Re = 100$.}
\end{figure}

\begin{figure}
 	\centering
 	\subfigure[]{
 	 		\label{cylindersidesides2biasCd}
 	 		\includegraphics[width=0.4 \textwidth]{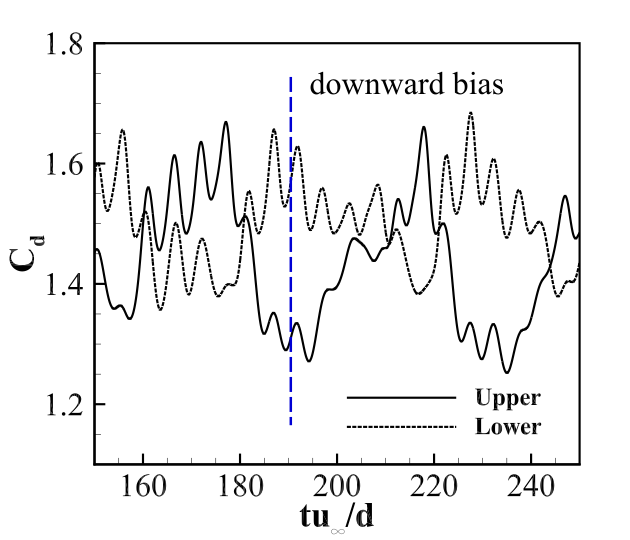}
 	 		\includegraphics[width=0.4 \textwidth]{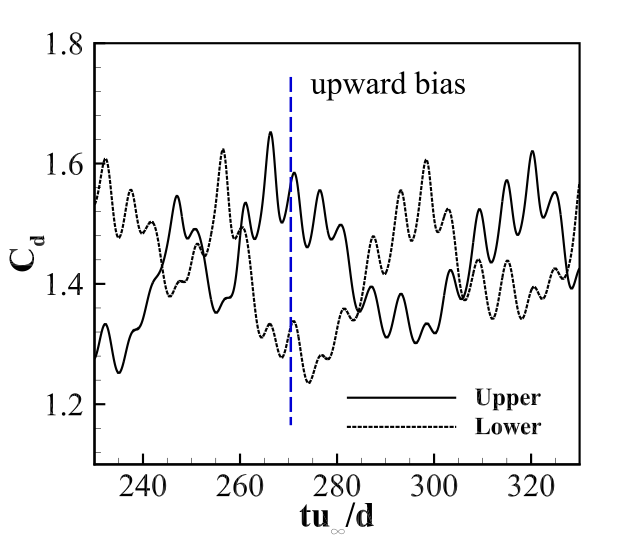}
 	 		}
     \subfigure[]{
   	 	 	\label{cylindersidesides2biasCl}
     	 	\includegraphics[width=0.4 \textwidth]{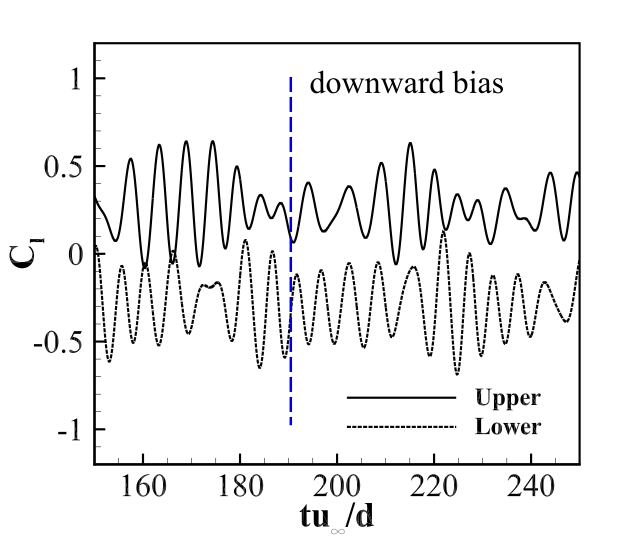}
     	 	\includegraphics[width=0.4 \textwidth]{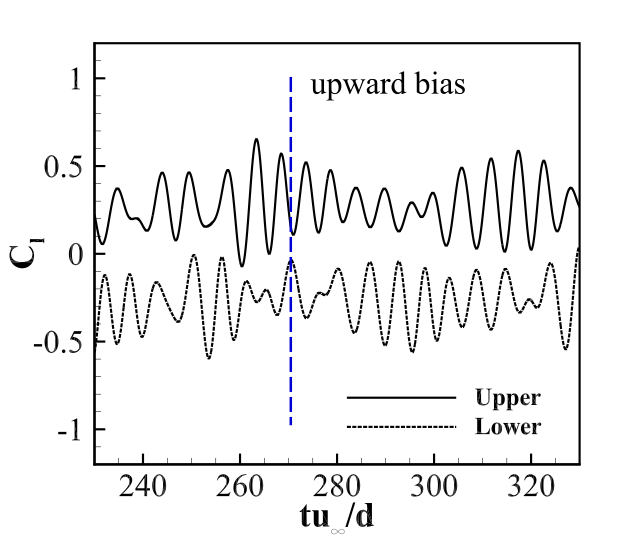}
     	 	}
 	\subfigure[]{
 			\label{cylindersidesides2streamlines}
 			\includegraphics[width=0.4 \textwidth]{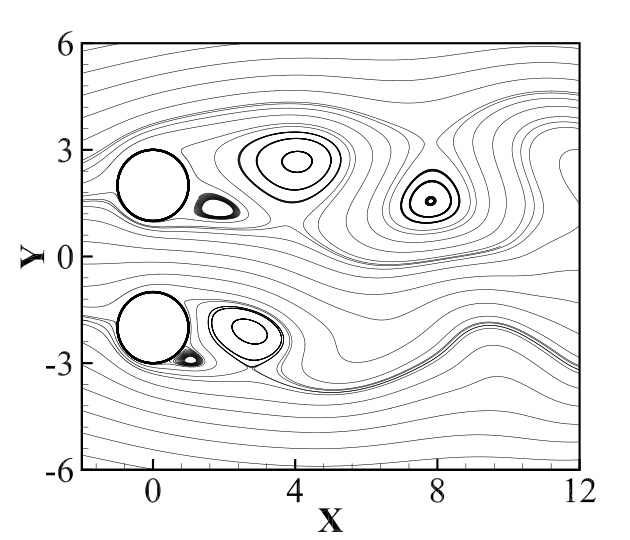}
 			\includegraphics[width=0.4 \textwidth]{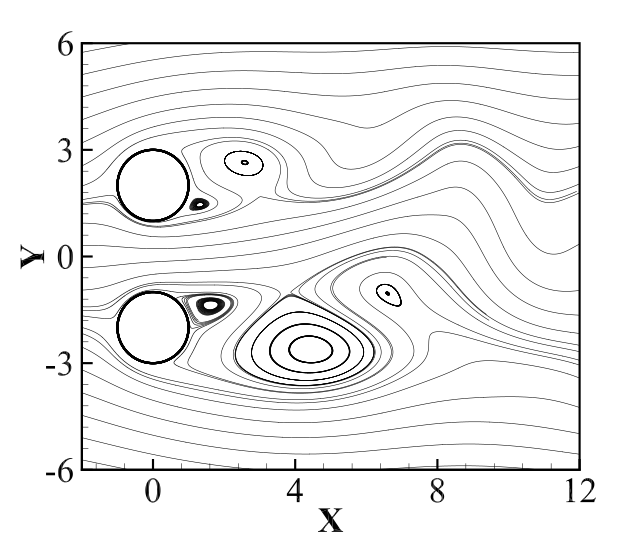}
 			}
     \subfigure[]{
 			\label{cylindersidesides2vorticity}
 			\includegraphics[width=0.4 \textwidth]{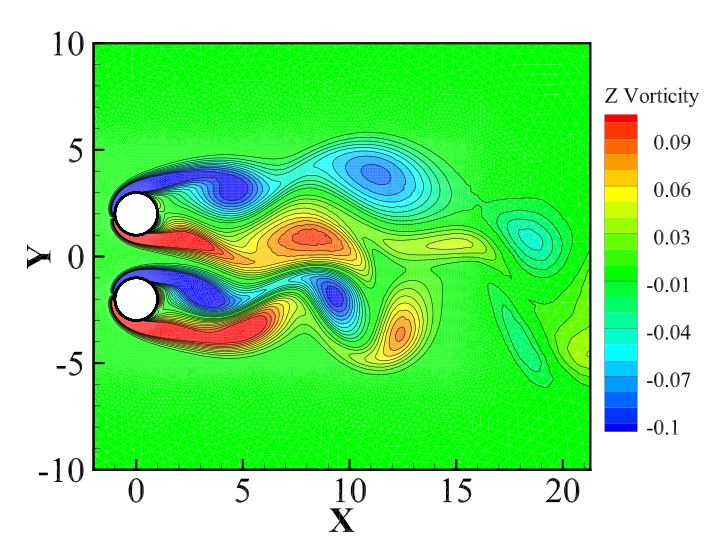}
 			\includegraphics[width=0.4 \textwidth]{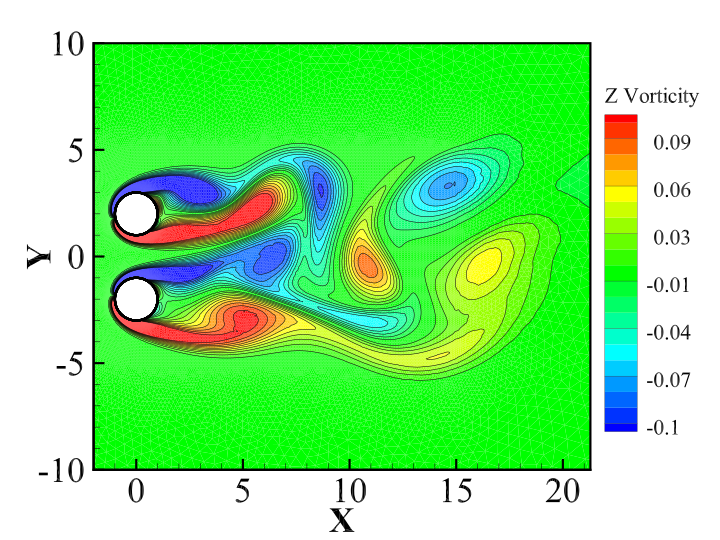}
 			}
 	\caption{\label{cylindersidesides2streamlinesvorticity} (a) Time evolutions of the drag coefficient, (b) the lift coefficient, (c) instantaneous streamlines and (d) instantaneous vorticity contours for gap flows downward (lift) and upward (right) biased between two cylinders arranged side-by-side with spacing $s = 2$ at $Re = 100$.}
\end{figure}

Compared with flow around single cylinder or two cylinders arranged in tandem, the simulating time that generate first shedding vortexes is much decline. Fig.~\ref{cylinderClCdsidesides2} shows irregular temporal variation of the lift coefficient $C_l$ and the drag coefficient $C_d$ at spacing $s = 2$. The same results can be found in Ref.~\cite{patil2012two,lee2009flow,liang2009high}. At this spacing, the gap flow between two cylinders is upward or downward biased and the shedding pattern of vortexes from two cylinders is asymmetric. For the downward biased, the lower cylinder will has higher drag coefficient and higher shedding frequency than the upper one. Besides, for the lift coefficient, the lower cylinder has a higher shedding frequency, about twice that of upper cylinder (see Fig.~\ref{cylindersidesides2streamlinesvorticity}). These results are in good agreement with many previous works and experimental observations\cite{chang1990interactive,bearman1973interaction}.

\begin{figure}
 	\centering
 	\subfigure{
 	 		\includegraphics[width=0.5 \textwidth]{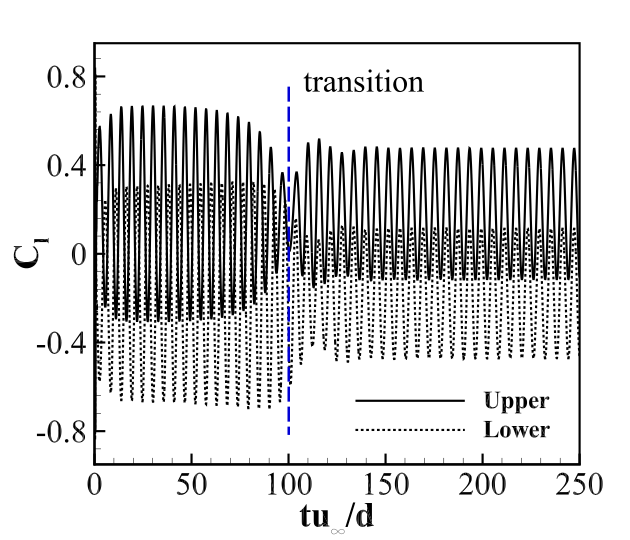}
 	 			}
 	\caption{\label{cylindersidesides2p5changephase} Transition of the lift coefficient from symmetric vortex shedding to antisymmetric vortex shedding for flow around two cylinders arranged side-by-side with spacing $s = 2.5$ at $Re = 100$.}
\end{figure}

\begin{figure}
 	\centering
 	\subfigure[]{
 			\label{cylindersidesides2p5Cla}
 			\includegraphics[width=0.45 \textwidth]{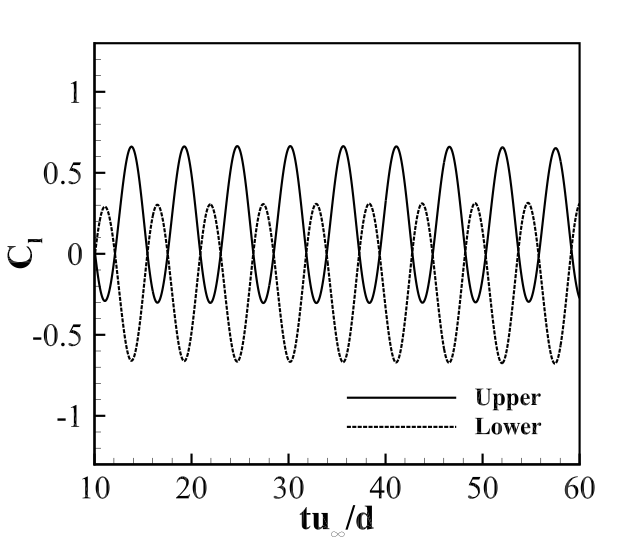}
 			}
    \subfigure[]{
    		\label{cylindersidesides2p5Cda}
   			\includegraphics[width=0.45 \textwidth]{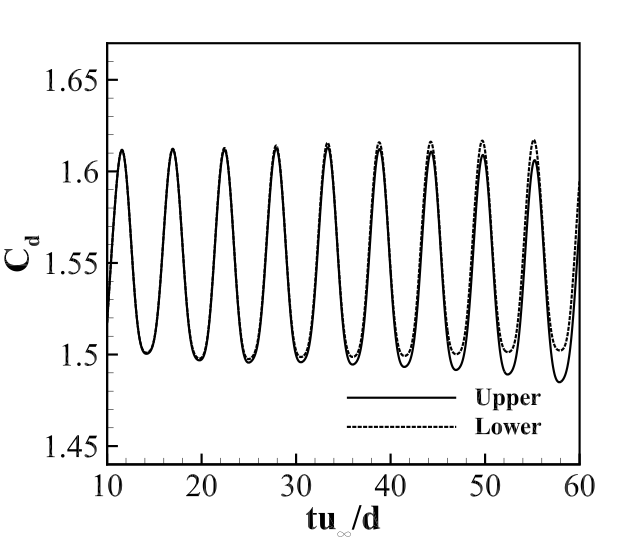}
     		}
 	\caption{\label{cylindersidesides2p5ClCda} Time evolutions of (a) the lift coefficient and (b) the drag coefficient at symmetric vortex shedding pattern for flow around two cylinders arranged side-by-side with spacing $s = 2.5$ and $Re = 100$.}
\end{figure}

\begin{figure}
 	\centering
 	\subfigure[]{
 			\label{cylinderside2p5symstream}
 			\includegraphics[width=0.4 \textwidth]{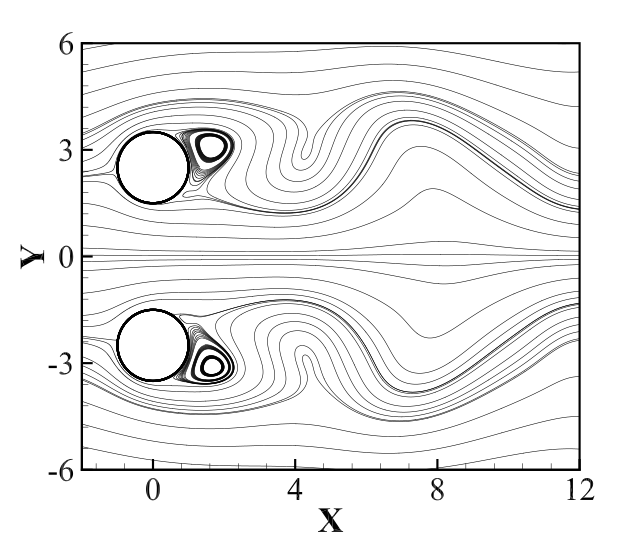}
 			}
    \subfigure[]{
    		\label{cylinderside2p5symvorticity}
   			\includegraphics[width=0.46 \textwidth]{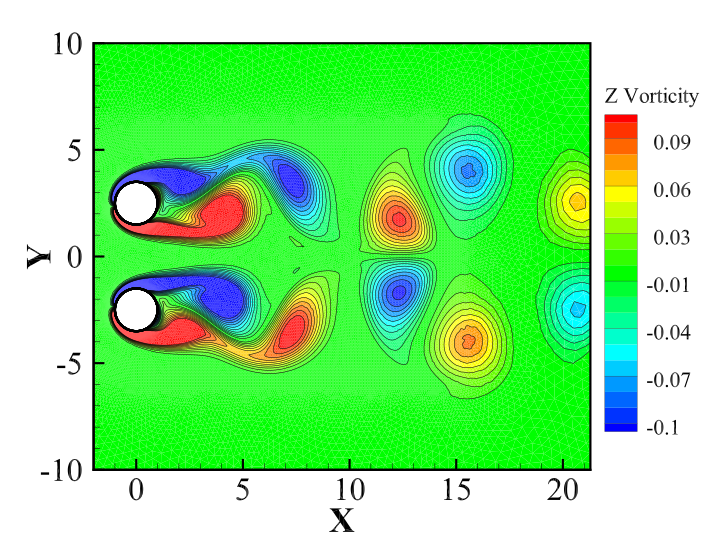}
     		}
 	\caption{\label{cylinderside2p5sym} (a) Instantaneous streamlines and (b) instantaneous vorticity contours of symmetric vortex shedding pattern for flow around two cylinders arranged side-by-side with spacing $s = 2.5$ at $Re = 100$.}
\end{figure}

\begin{figure}
 	\centering
 	\subfigure[]{
 			\label{cylindersidesides2p5Cl}
 			\includegraphics[width=0.45 \textwidth]{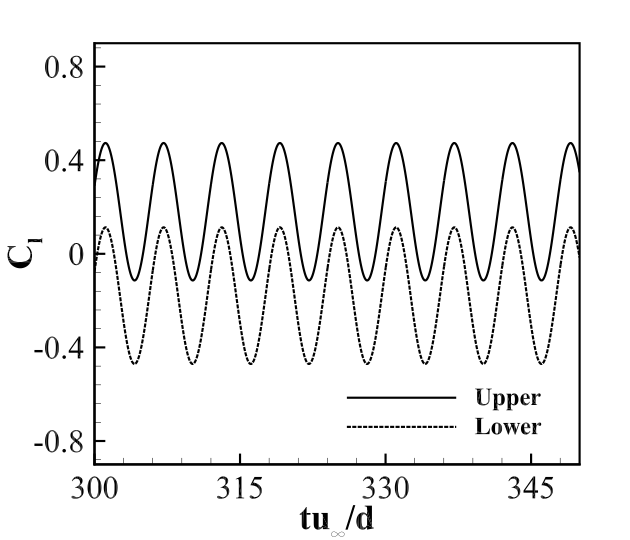}
 			}
    \subfigure[]{
    		\label{cylindersidesides2p5Cd}
   			\includegraphics[width=0.45 \textwidth]{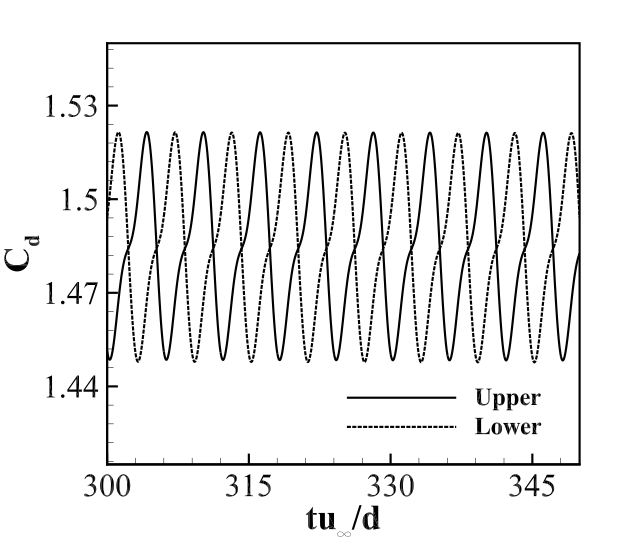}
     		}
 	\caption{\label{cylindersidesides2p5ClCd} Time evolutions of (a) the lift coefficient and (b) the drag coefficient at antisymmetric vortex shedding pattern for flow around two cylinders arranged side-by-side with spacing $s = 2.5$ and $Re = 100$.}
\end{figure}

\begin{figure}
 	\centering
     \subfigure{
   			\includegraphics[width=0.45 \textwidth]{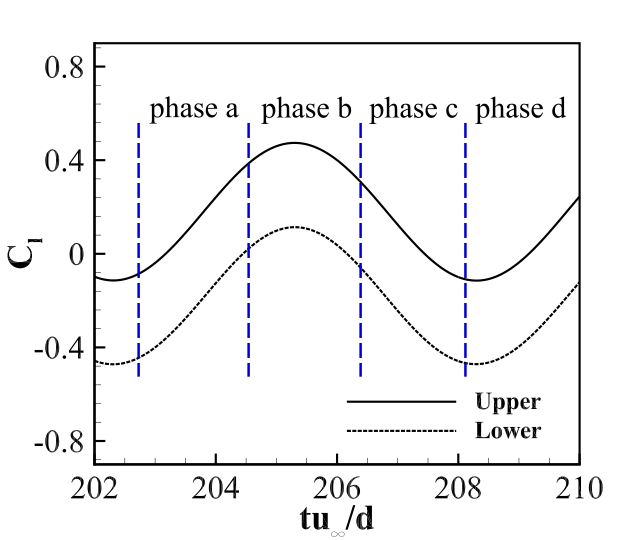}
     		}
 	\caption{\label{cylindersidesides2p5phase} Definition of four-phase at antisymmetric vortex shedding pattern for flow around two cylinders arranged side-by-side with $s = 2.5$ and $Re = 100$.}
\end{figure}

\begin{figure}
 	\centering
     \subfigure[Phase a]{
   			\includegraphics[width=0.4 \textwidth]{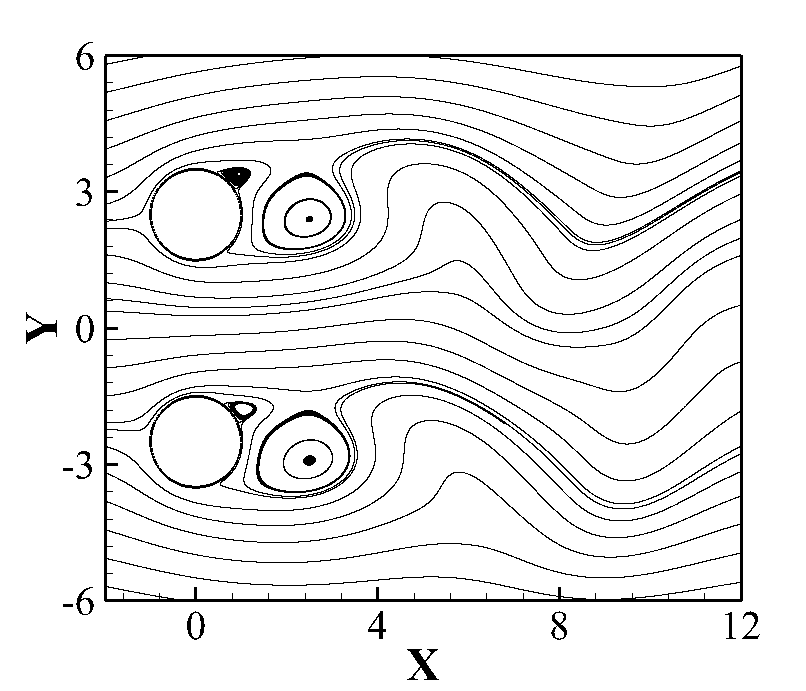}
   			\includegraphics[width=0.46 \textwidth]{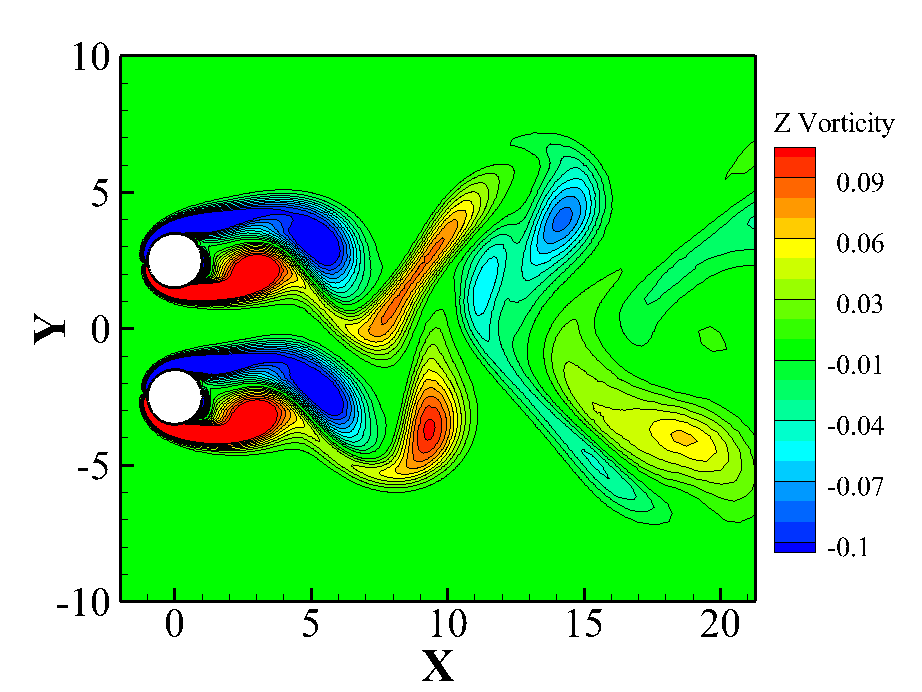}
     		}
     \subfigure[Phase b]{
        	\includegraphics[width=0.4 \textwidth]{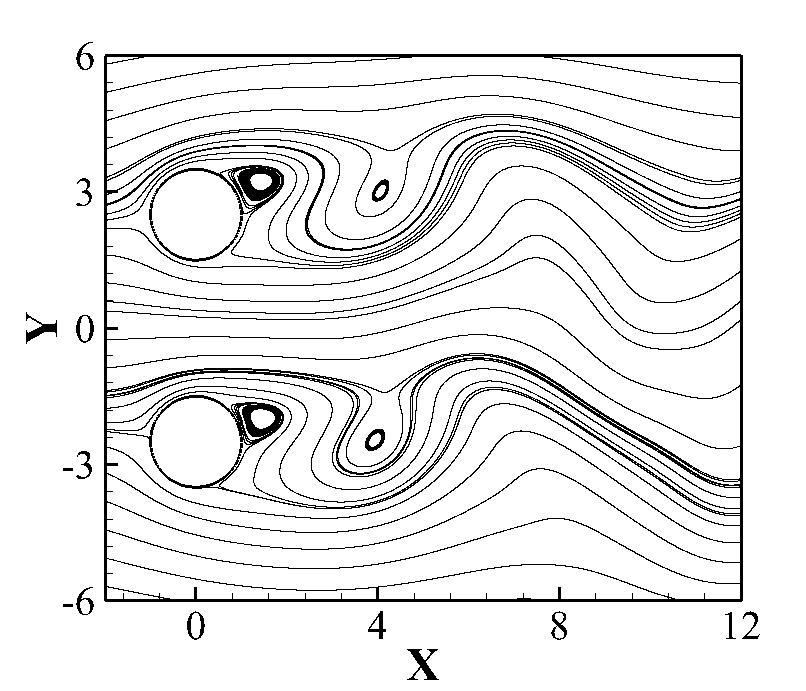}
       		\includegraphics[width=0.46 \textwidth]{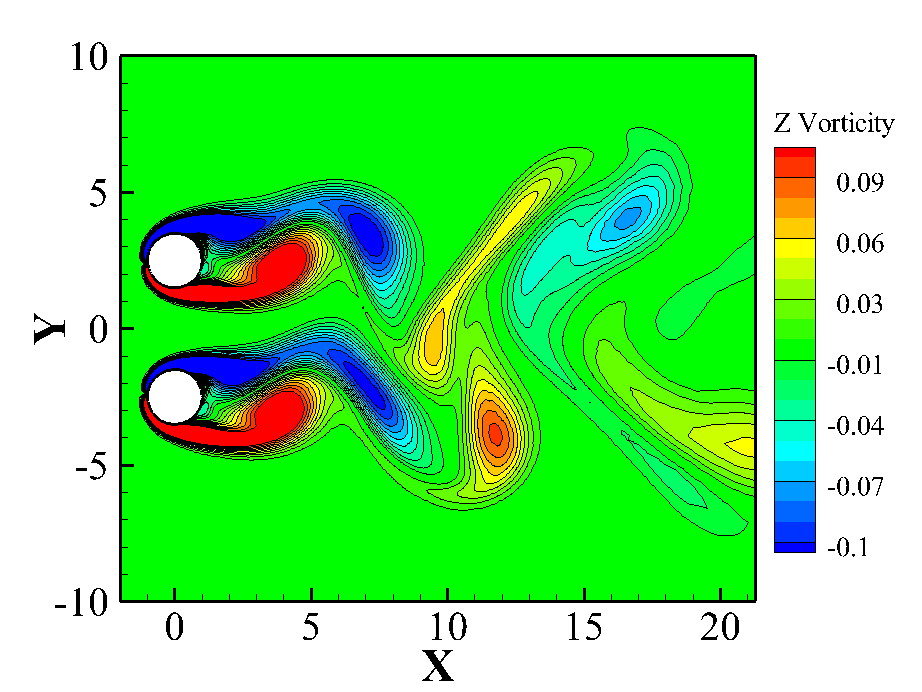}
          		}
     \subfigure[Phase c]{
           	\includegraphics[width=0.4 \textwidth]{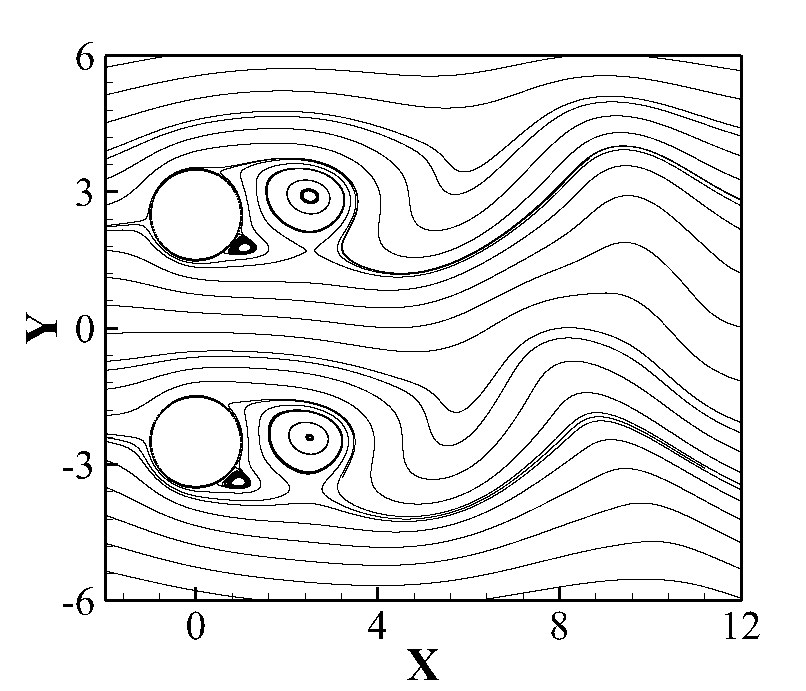}
       		\includegraphics[width=0.46 \textwidth]{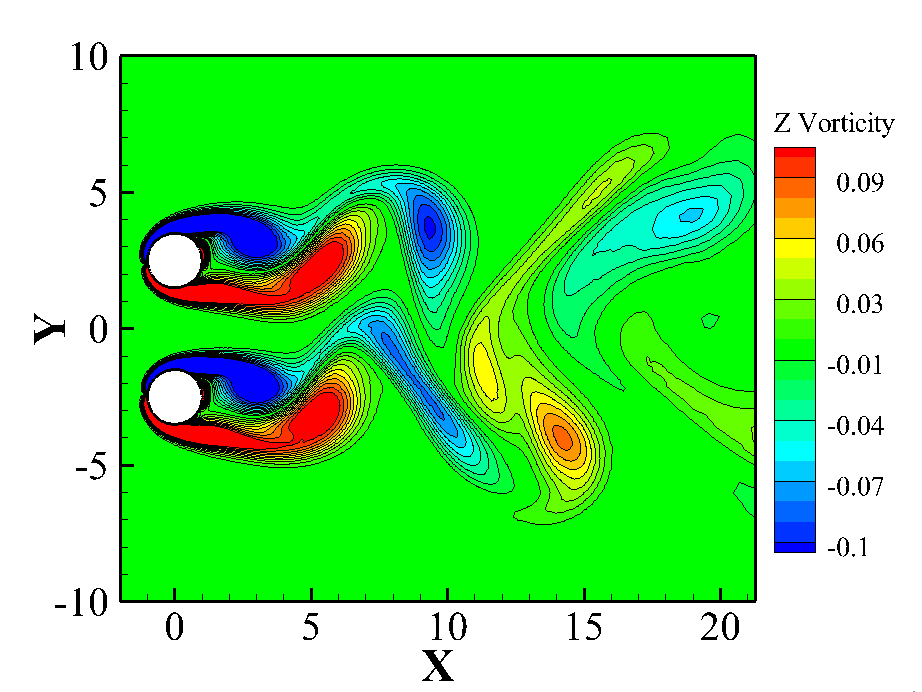}
               	}
     \subfigure[Phase d]{
           \includegraphics[width=0.4 \textwidth]{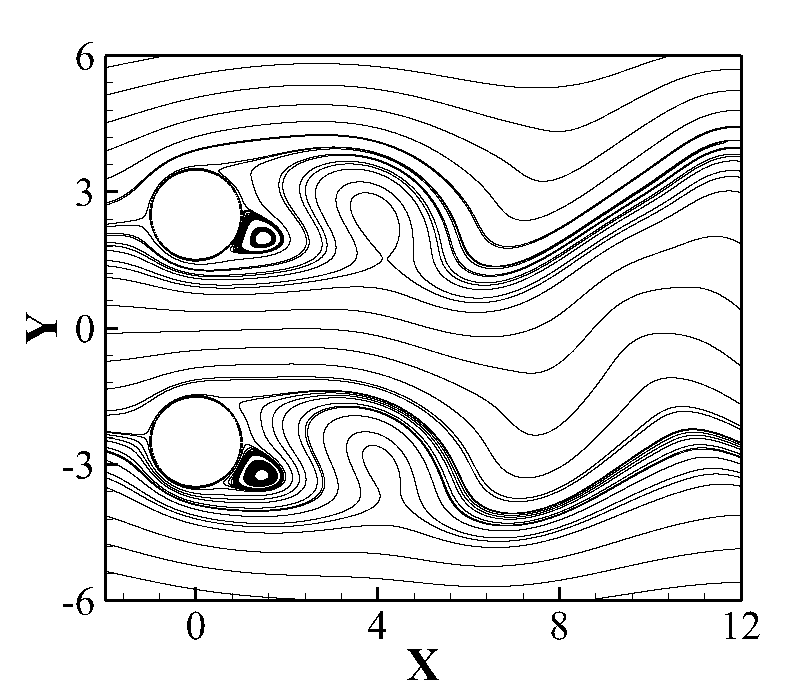}
           \includegraphics[width=0.46 \textwidth]{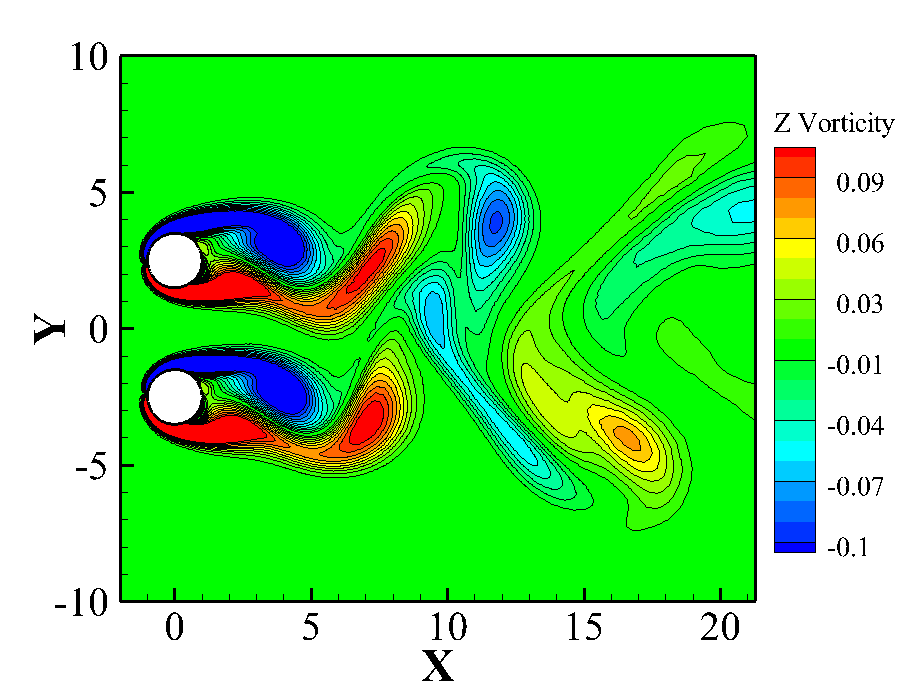}
                }
 	\caption{\label{cylinderside2p5stream} Instantaneous streamlines and vorticity contours at different phases for flow around two cylinders arranged side-by-side with spacing $s = 2.5$ and $Re = 100$.}
\end{figure}

Fig.~\ref{cylindersidesides2p5changephase} shows the time history of lift coefficient that change from anti-phase to in-phase, the corresponding vortex shedding pattern is changed from symmetric to antisymmetric. At the symmetric shedding pattern, the $C_l$ is anti-phase and $C_d$ is in-phase (see Fig.~\ref{cylindersidesides2p5ClCda}), and the dividing streamlines are straight. The vortexes shedding from the cylinders do not merger and can maintain there forms for a relatively longer time (see Fig.~\ref{cylinderside2p5sym}). When antisymmetric shedding pattern appearing, the $C_l$ will change into in-phase and $C_d$ is anti-phase (see Fig.~\ref{cylindersidesides2p5ClCd}). The same definition of phase of $C_l$ in Ref.~\cite{liang2009high} is presented in Fig.~\ref{cylindersidesides2p5phase}, and corresponding streamlines and vorticity contours at different phase can be found in Fig.~\ref{cylinderside2p5stream}. In the evolution from phase a to phase d, a pair of vortexes will alternating generate at the top of cylinders or at the bottom of cylinders, then flow away from the surfaces of cylinders. At antisymmetric shedding pattern, the shedding vortexes can not maintain long time and will merge each others. The drag coefficient and lift coefficient are compared with other works in Table \ref{tab:cylindersideside}.  It is shown that our results are in reasonable range.

\begin{table}
    \centering
	\caption{\label{tab:cylindersideside} The drag coefficient and lift coefficient of upper cylinder for flow around two cylinders arranged side-by-side.}
	\begin{tabular}{ccccc}
	  \toprule
	    \multicolumn{1}{c}{Reference} & \multicolumn{1}{c}{$s$} & \multicolumn{1}{c}{$\overline{C}_d$} &   \multicolumn{1}{c}{$\overline{C}_l$} &  \multicolumn{1}{c}{$\tilde{C}_l$} \\
	  \midrule
	     Lee et al.\cite{lee2009flow}       & 2   & 1.4390 & 0.2727 & 0.1574 \\
	     Kang\cite{kang2003characteristics} & 2   & 1.4429 & 0.2761 & 0.1686 \\
	     Patil et al.\cite{patil2012two}    & 2   & 1.4074 & 0.2619 & 0.1257 \\
	     Present                            & 2   & 1.4532 & 0.2711 & 0.1627 \\
	  \midrule
	     Lee et al.\cite{lee2009flow}       & 2.5 & 1.4538 & 0.1848 & 0.1976 \\
	     Kang\cite{kang2003characteristics} & 2.5 & 1.4272 & 0.1780 & 0.1859 \\
	     Present                            & 2.5 & 1.4842 & 0.1817 & 0.2074 \\
	  \bottomrule
	\end{tabular}
\end{table}

\section{Conclusions}\label{Sec:conclusions}
In this study, the original finite volume LBM presented by Peng et al.\cite{peng1998lattice}, Ubertini et al.\cite{ubertini2004recent}, and Stiebler et al.\cite{stiebler2006upwind} are further investigated. For the reconstruction of distribution function at the cell interface, the LSLR upwind scheme based on the cell-vertex finite volume method is extended to the cell-center finite volume method. Different boundary conditions are illustrated in detail and can be easy implemented. From several test cases, the present method can apply to both steady and unsteady laminar flow. The defect of FVLBM is much lower computational efficiency compared with standard LBM and need further study to alleviate this problem. The main feature of present FVLBM can conclude as
\begin{description}
    \item[(1)] the hybrid mesh used in test cases of flow around circular cylinder shows the great flexibility for treatment the complex geometries.

    \item[(2)] both spatial and temporal discretization can achieve second-order accuracy.

    \item[(3)] for unsteady flow, the present method can have good results compared with other numerical methods.

    \item[(4)] easy to implement parallel computing as only one layer of ghost cell at the interfaces of partitions is needed.

    \item[(5)] can be easy extended to three-dimensional scheme.
\end{description}

\section*{Acknowledgements}
This work is supported by National Natural Science Foundation of China (11472219), 111 project of China (B17037) and the Discovery Grant of the Natural Sciences and Engineering Research Council (NSERC) of Canada.

\clearpage
\section*{References}
\bibliography{ref}

\end{document}